\documentclass[usenatbib]{mn2e}
\usepackage{graphicx}
\usepackage{color}
\def\deg{\ifmmode^\circ\else$^\circ$\fi}

\title[High mass-loss AGB stars detected by MSX in the Galactic bulge]
{High mass-loss AGB stars detected by MSX in the ``intermediate" and
``outer" Galactic bulge}

\author[D.K. Ojha et al.]{D.K. Ojha,$^{1}$\thanks{E-mail: ojha@tifr.res.in},
A. Tej$^{1}$, M. Schultheis$^{2}$, A. Omont$^{3}$ and F. Schuller$^{4}$\\
$^{1}$Tata Institute of Fundamental Research, Homi Bhabha Road, Colaba,
Mumbai (Bombay) - 400 005, India
\\
$^{2}$Observatoire de Besan\c con, BP 1615, F-25010 Besan\c con Cedex,
France
\\
$^{3}$Institut d'Astrophysique de Paris, CNRS, 98bis Bd Arago, F-75014
Paris, France
\\
$^{4}$Max Planck Institut fur Radioastronomie, D-53121 Bonn, Germany
}

\begin{document}

\date{}
\pagerange{\pageref{firstpage}--\pageref{lastpage}} \pubyear{2007}

\maketitle

\label{firstpage}

\begin{abstract}

We present a study of MSX point sources in the Galactic bulge  
($|$$l$$|$ $<$ 3$^\circ$, 1$^\circ$ $<$ $|$$b$$|$ $<$ 5$^\circ$),
observed at A, C, D and E-band (8 to 21 $\mu$m), with a total area 
$\sim$ 48 deg$^2$ and more than 7000 detected sources in the MSX D-band 
(15 $\mu$m). We discuss the nature of the MSX sources (mostly AGB stars),
their luminosities, the interstellar extinction, the mass-loss rate 
distribution and the total mass-loss rate in the bulge.
The mid-infrared data of MSX point sources have been 
combined with the near-infrared ($J$, $H$ and $K_{\rm s}$) data of 2MASS survey. 
The cross-identification was restricted to $K_{\rm s}$-band detected sources with 
$K_{\rm s}$ $\le$ 11 mag. However, for those bright MSX D-band sources 
([D] $<$ 4.0 mag), which do not satisfy this criteria,
we have set no $K_{\rm s}$-band magnitude cut off.
The bolometric magnitudes and the corresponding luminosities of the
MSX sources were derived by fitting blackbody curves.
The relation between $\rm \dot{M}$ and ($K_{\rm s}$-[15])$_{\rm 0}$ was used to derive 
the mass-loss rate of each MSX source in the bulge fields.
Except for very few post-AGB stars, planetary nebulae and
OH/IR stars, a large fraction of the detected sources at 15 $\mu$m 
(MSX D-band) are AGB stars well above the RGB tip. A number of them show 
an excess in ([A]-[D])$_0$ and ($K_{\rm s}$-[D])$_0$ colours, characteristic 
of mass-loss. These colours, especially ($K_{\rm s}$-[D])$_0$,
enable estimation of the mass-loss rates ($\rm \dot{M}$) of the sources
in the bulge fields which range from
$\rm 10^{-7}$ to $\rm 10^{-4}$ $\rm M_{\odot}\, yr^{-1}$.
Taking into consideration the completeness of the mass-loss rate bins, 
we find that the contribution to the integrated mass-loss  
is probably dominated by mass-loss rates larger than $\rm 3 \times 10^{-7}$ 
$\rm M_{\odot}\, yr^{-1}$ and is about 
$\rm 1.96\times10^{-4}\, M_{\odot}\, yr ^{-1}\, deg^{-2}$ in the 
``intermediate" and ``outer" bulge fields of sources with mass-loss rates, 
$\rm \dot{M} > 3\times10^{-7}\, M_{\odot}\, 
yr^{-1}$. The corresponding integrated mass-loss rate per unit stellar mass is 
$\rm 0.48\times10^{-11}\, yr^{-1}$.
Apart from this, the various mid- and near-infrared colour-colour and
colour-magnitude
diagrams are discussed in the paper to study the nature of the stellar
population in the MSX bulge fields. 
\end{abstract}

\begin{keywords}
stars: AGB and post-AGB -- stars: circumstellar dust - stars: 
mass-loss - dust: extinction - infrared: stars
\end{keywords}

\section{Introduction}

Study of the stellar population of the Galactic bulge fields is of prime importance and 
plays a crucial role in understanding the formation and evolution of the Galaxy.
The high luminosity Asymptotic Giant Branch (AGB) stars are ideal tracers of 
stellar population in regions of high extinction such as the ``intermediate"
 and ``outer" Galactic bulge. Intense mass-loss 
($\ga \rm 10^{-6}\, M_{\odot}\, yr^{-1}$)
phase has been identified with stars 
evolving along the AGB phase. Hence, these sources are
enshrouded with circumstellar envelopes 
of dust and gas. The low effective temperatures and thermal emission from 
warm dust 
make these stars bright in the infrared. Deep and large area infrared surveys offer 
a unique view of the stellar population towards the inner Galaxy as high 
interstellar extinction hinders the study at optical wavelengths. A combination 
of mid- and near-infrared data is essential to sample the high 
mass-loss AGB population. Mid-infrared data are more sensitive to the infrared 
excess which is a consequence of mass-loss in the AGB stars. With the 
availability of infrared data from surveys like DENIS (Epchtein et al. 1994), 
2MASS (Beichman et al. 1998; Skrutskie et al. 2006), ISOGAL (Omont et al. 2003) and 
MSX (Price et al. 2001), there have been a large number of studies on the AGB 
population of the Galactic bulge fields (e.g. Glass et al. 1999; 
Schultheis \& Glass 2001; Glass \& Schultheis 2002; Groenewegen \& 
Blommaert 2005). In a recent study, 
Ojha et al. (2003) have combined the mid- and near-infrared photometry from 
ISOGAL, DENIS and 2MASS to study the nature of ISOGAL sources in the 
``intermediate" Galactic bulge and discuss their mass-loss rates.
Further, the data
from GLIMPSE II survey (Ed Churchwell; private communication) 
will provide an excellent 
opportunity to study the AGB stars in the inner bulge region.
However, the sample has to be restricted to high extinction regions
owing to saturation effects of the survey. In the present study we
have not included GLIMPSE II data as it does not cover the entire
``intermediate" and ``outer" bulge regions selected in this paper. The 
stellar population
study of the inner bulge with GLIMPSE II data will be presented in 
a future paper.  The AGB stars contribute more than 70\% towards the 
enrichment of the dust component of the ISM in the solar
neighbourhood (Sedlmayer 1994) and hence it is important to study their 
mass-loss in different parts of the Galaxy. As discussed in Ojha et al. (2003), 
the total mass returned to the ISM is dominated by mass-loss rates greater than 
$\rm 10^{-6}\, M_{\odot}\, yr^{-1}$ and hence the detection of the entire population 
of the high mass-loss stars becomes important for determination of the 
total mass returned to the ISM and probably the total mass of the bulge.    

In this paper, we report the study of the MSX point sources in the 
``intermediate" and ``outer" Galactic bulge fields  
($|$ $l$ $|$ $<$ 3$^\circ$, 1$^\circ$ $<$ $|$$b$$|$ $<$ 5$^\circ$)
with a total area $\sim$ 48 deg$^2$ and more than 7000 detected sources 
in the MSX D-band (15 $\mu$m). Here, the division into the
two aforementioned bulge fields is based on the Galactic latitude. 
The MSX bulge fields in the
Galactic latitude bins 1$^\circ$ $<$ $|$$b$$|$ $<$ 2$^\circ$ and 
 2$^\circ$ $<$ $|$$b$$|$ $<$ 5$^\circ$ are defined as the ``intermediate" and
the ``outer" bulge regions, respectively. 
We restrict our study to the ``intermediate" and 
``outer" Galactic bulge because the extinction here is much less and 
more homogeneous as compared to the inner bulge.
The density of the sources in ``intermediate" and
``outer" bulge also reduce the number of spurious associations to the minimum. 
As shown in Price et al. (2001), 
uniform sky coverage in the MSX bands extends for only the 
northern latitudes upto $b = 6^\circ$
and the higher latitudes (outer Galaxy)
are very sparsely sampled. We have restricted the latitude
selection such that the fields covered by us are free from the non-uniform sky
coverage beyond $|$$b$$|$ $>$ 5$^\circ$.
We have combined the mid-infrared data of the 
MSX point sources with the near-infrared 2MASS data
to determine their nature and the interstellar extinction. 
Most of the sources are AGB stars well above the RGB tip with high mass-loss. 
We have determined the luminosities and mass-loss rates of the stellar
population in the MSX bulge fields. 

The outline of the paper is as follows : In \S 2, we present the 
MSX and 2MASS observations and describe the cross-identification
between MSX and 2MASS sources.
In \S 3, we discuss the determination of interstellar extinction in
the line of sight of the bulge fields using the isochrone
fitting method.
The ISOGAL and DENIS associations of the MSX sources in the bulge fields
are discussed in \S 4.
\S 5 presents the derivation of bolometric magnitude ($M_{bol}$) and
luminosity for each star in the bulge fields. 
In \S 6, we discuss the derivation of mass-loss rates based
on ($K_{\rm s}$-[D])$_0$ colour and the estimation of the total mass-loss
rate in the ``intermediate" and ``outer" bulge.
In \S 7, we present the nature of the MSX sources from the various
colour-colour and colour-magnitude diagrams.
We summarize our results in \S 8.

\section{Observations and cross-correlation of MSX and 2MASS sources}

We have used the deep MSX Point Source Catalog Version 2.3 
(Egan et al. 2003) in this paper.
The catalogue (MSXPSC V2.3) lists the sources detected in MSX
mid-infrared bands A, C, D and E with $\lambda(\Delta\lambda)$ corresponding
to 8.28(3.36), 12.13(1.72), 14.65(2.23) and 21.34(6.24) $\mu$m, respectively.
MSXPSC V2.3 has several improvements over the initial
published catalogue, MSXPSC Version 1.2 (Egan et al. 1999). 
In this latest version, the photometry is based on co-added image plates, as opposed
to single-scan data, which results in improved sensitivity and hence
reliability in the fluxes. Comparison with Tycho-2 positions indicates that
the positional accuracy, $\sigma$ = 2\arcsec~at 8 $\mu$m,  
of the new catalogue 
is better as compared to MSXPSC V1.2. Also, MSXPSC V2.3 is 100\%
complete in the A-band (8\,$\mu$m) and
approaches 100\% completeness in the other three bands at the
survey sensitivity limit of the MSX image data ($\sim$ 0.1 Jy in
A-band, which is more sensitive than the other bands by a factor of
$\sim$ 10)(Egan et al. 2003).

\begin{figure*}
\centering
\resizebox{\hsize}{!}{\includegraphics{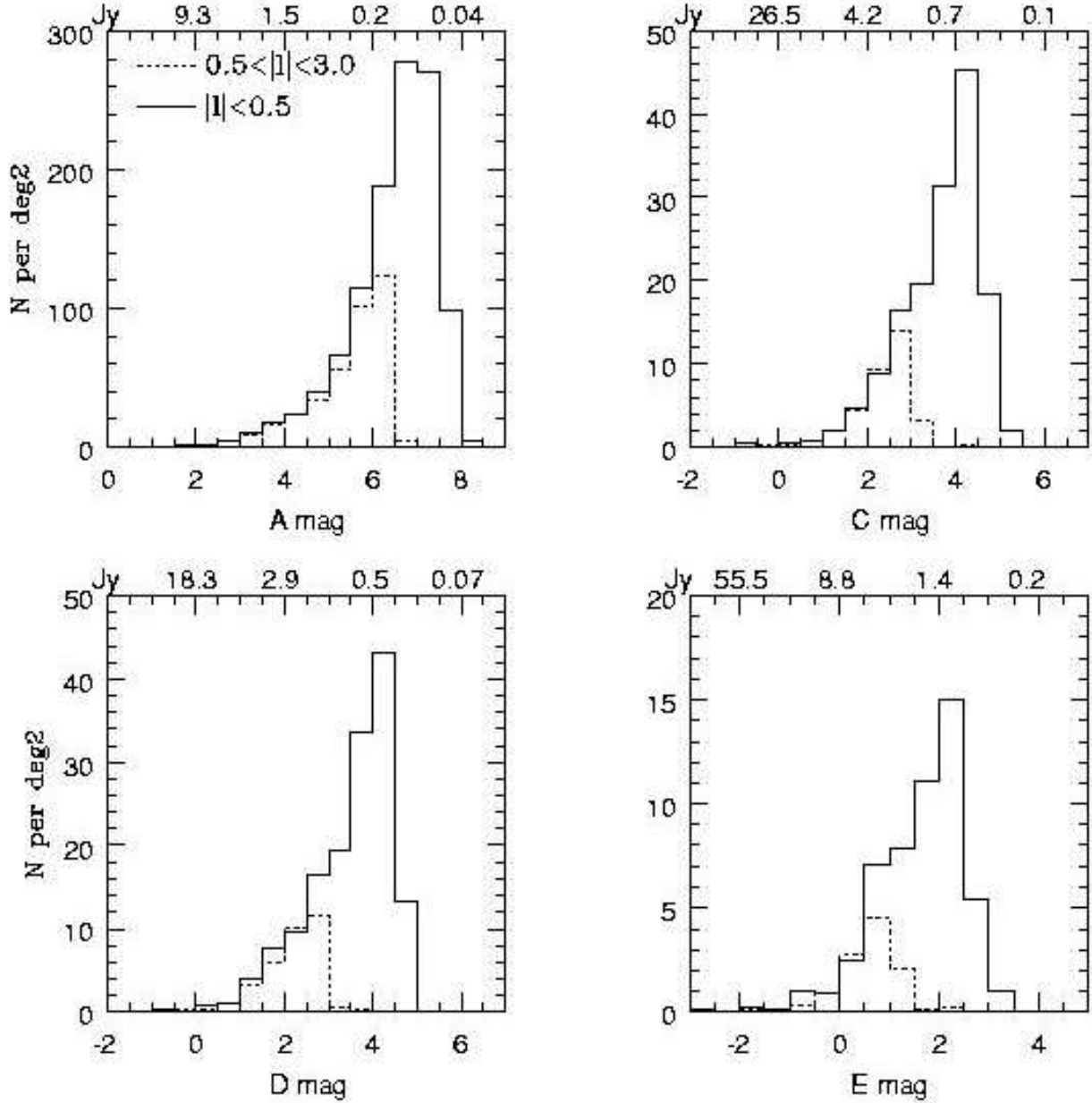}}
\caption{A, C, D and E-band source distributions in half magnitude bins
for good quality (quality flags 3 \& 4) MSX data. 
Magnitudes are converted into fluxes using the zero magnitude flux given by 
Egan et al. (1999). The solid line denotes the MSX sources for
the ``inner"
zone ($|l| < 0.5^\circ$) and the dotted line represents the sources for 
the ``outer" zone ($0.5^{\circ} < |l| < 3.0^\circ$) in the bulge.   
}
\label{FIG1}
\end{figure*}

The total number of sources detected in our fields are
29709, 8125, 7390, and 3380 in the MSX A-, C-, D- and E-bands, respectively.
In the study that follows, we have included only
good quality MSX data for which the flux quality flags are 3 and 4
(which implies S/N $\ga$ 7).
Taking the quality flags into account, the total number of 
good quality sources in our fields are 24035, 2571, 2550
and 866 in the four bands, respectively. The corresponding average source
densities are $\sim$ 5.0$\times$10$^2$ deg$^{-2}$,
$\sim$ 0.5$\times$10$^2$ deg$^{-2}$, $\sim$
0.5$\times$10$^2$ deg$^{-2}$
and $\sim$ 0.2$\times$10$^2$ deg$^{-2}$ in the four bands, respectively.

The MSX catalogue is more sensitive (by a factor of $\sim$ 4 
in the A-band) towards the inner Galactic longitudes as compared to the
outer longitudes. This is due to the fact that additional maps of regions
at $l=0$ were co-added to the survey data (Sean Carey; private communication). 
These additional maps are $\sim$ 4 times more sensitive. 
This is evident from the inspection of the data sets. 
Hence, for clarity we have
divided the catalogue into two sets for our study: ``inner" 
zone ($|l| < 0.5^{\circ}$) and ``outer" zone ($0.5^{\circ} < |l| <
3.0^{\circ}$). 
The magnitude histograms of the two zones in the MSX A-, C-, D- and E-bands 
are displayed in Fig.\,\,\ref{FIG1}. The histograms show the
total source counts of the good quality sources.
The 50\% completeness limits (as determined from the flux histograms)
of the ``inner" zone are 0.08, 0.40, 0.32 
and 1.00 Jy for A-, C-, D- and E-bands, respectively. For the
``outer" zone, the 50\% completeness limits correspond to
0.16, 1.58, 1.26 and 3.98 Jy for A-, C-, D- and E-bands, respectively. 

From the ISOGAL survey, it is seen that a number of
the ISOGAL 15\,$\mu$m sources are AGB stars in the Galactic bulge and
the central disk with evidence of mass-loss in majority of the cases
(Omont et al. 2003). The mass-loss in AGB stars may be characterized by their
15\,$\mu$m excess (Omont et al. 1999; Ortiz et al. 2002; 
Ojha et al. 2003). The MSX D-band
is close to this wavelength and hence in the study that follows, our prime
focus lies on the sources detected in the D-band. 
The proportions of D-band sources associated with A, C, and
E-band sources are $\sim$ 100\%, $\sim$ 91\% and $\sim$ 33\%,
respectively.

The near-infrared data used in this paper have been obtained from the 
2MASS Point Source Catalogue (2MASSPSC) in 
the three bands, $J$ (1.25 $\mu$m), $H$ (1.65 $\mu$m) and 
$K_{\rm s}$ (2.17 $\mu$m). The full sky 2MASS catalogue covers the entire MSX bulge fields. 
The 2MASS PSC is $>$99 \% complete in the absence of confusion at the 10$\sigma$
sensitivity limit of 15.8, 15.1 and 14.3 mag in the $J$-, $H$- and
$K_{\rm s}$-band, 
respectively (Skrutskie et al. 2006). However, the turnover in the source counts in the 
Galactic plane field occurs nearly 1 mag brighter, because of the effects of confusion noise on 
the detection thresholds. The primary areas of
confusion are (1) longitudes $\pm$75 deg from the Galactic center and
latitudes 1 deg from the Galactic plane and (2) within an approximately 
5 deg radius of the Galactic center (Skrutskie et al. 2006).

\begin{figure*}
\includegraphics[width=7.3cm]{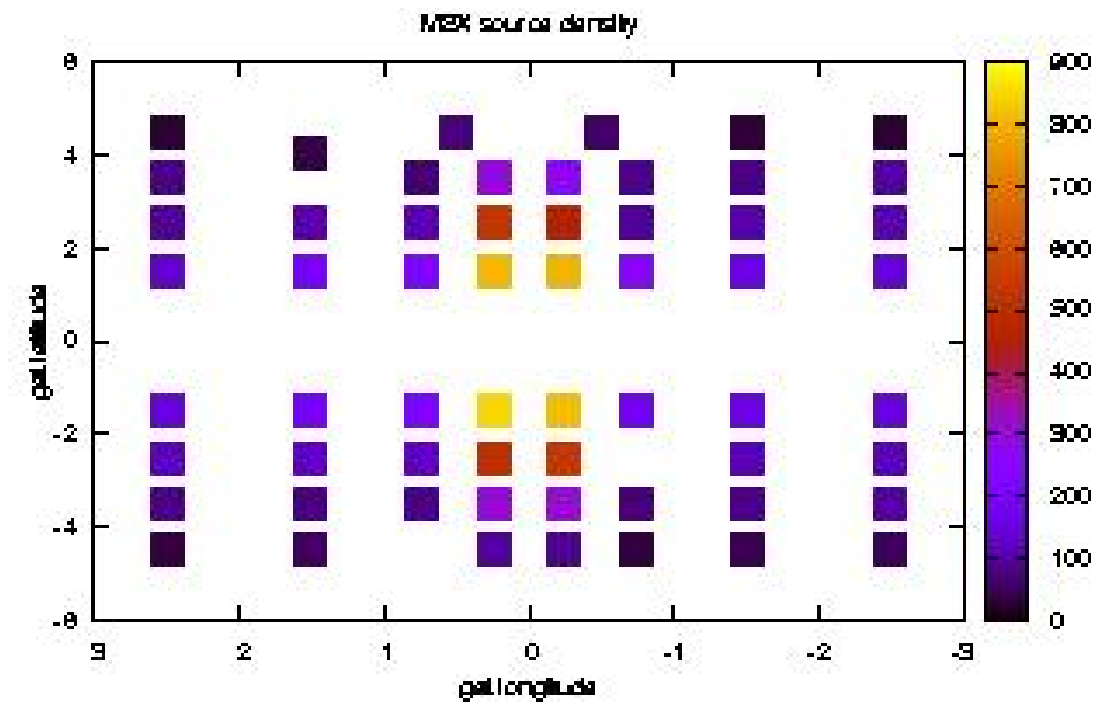}
\qquad
\includegraphics[width=7.3cm]{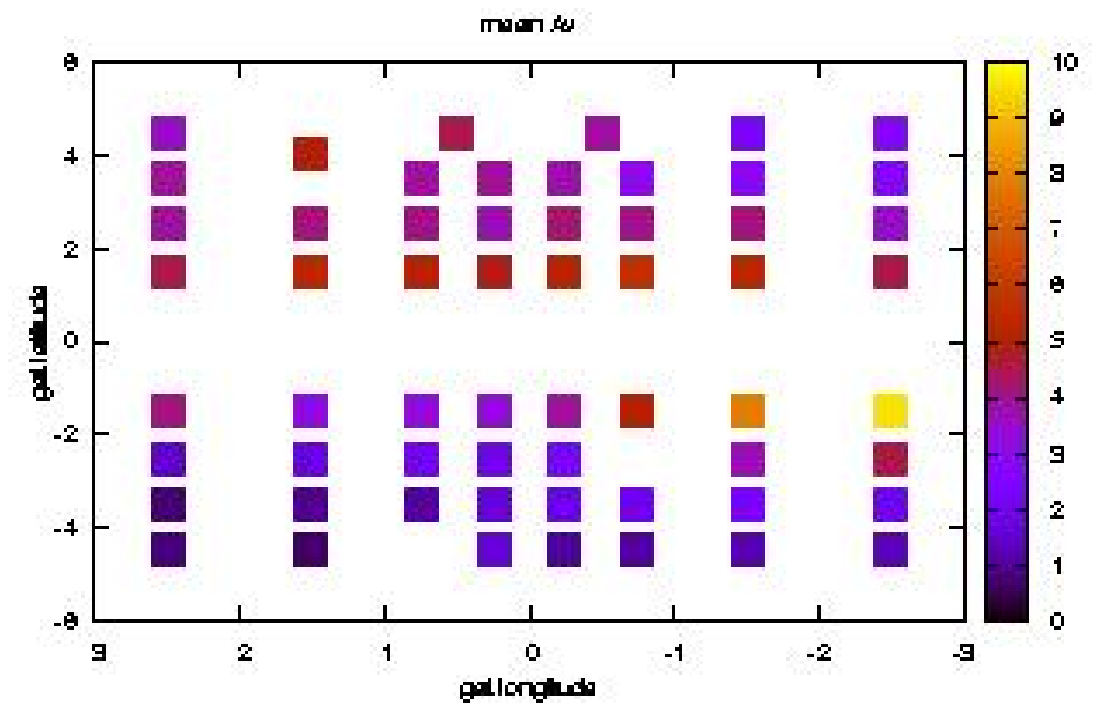}
\includegraphics[width=7.3cm]{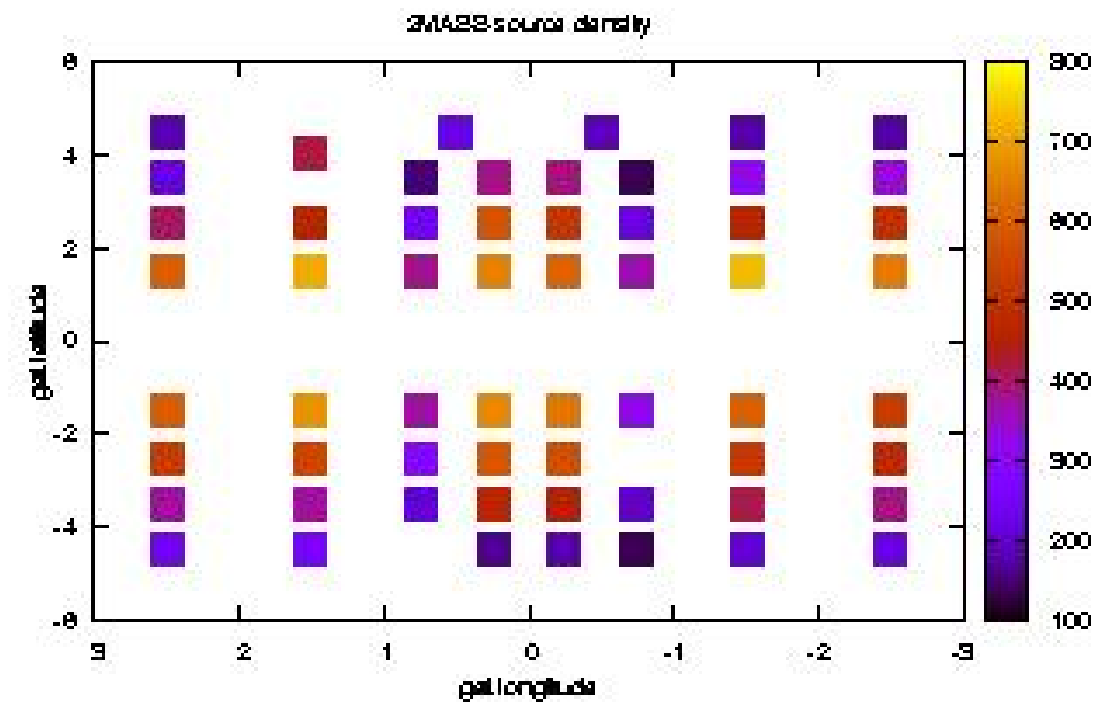}
\caption{The figure shows the 3-D plots of the MSX source density, the mean
A$_{V}$ and the total number of 2MASS sources ($K_{\rm s}$
$\la$ 11.0 mag) in the various subfields of the bulge as a function of the 
Galactic longitude and latitude. This information in tabular form
is presented in Table 1 which is only available in the electronic form 
via the VizieR Service at the CDS.}
\label{FIG2}
\end{figure*}

A cross-correlation algorithm similar to the one used by
Schuller et al. (2003), to associate the ISOGAL 7 and
15\,$\mu$m sources
with DENIS sources, has been used to search for 2MASS
counterparts of the MSX point sources in the bulge fields. 
However, to restrict the number of spurious detections at 
the fainter end, we set our
magnitude cut at conservative brighter level and retain
only those 2MASS sources with $K_{\rm s}$ $\la$ 11 mag for
cross-correlating with the MSX catalogue.
This 2MASS magnitude cut ($K_{\rm s} \le$ 11 mag) corresponds to
an average density of $K_{\rm s}$-band sources $n$ $\sim$ 22 300
per deg$^2$.
However, the density of $K_{\rm s} \le$ 11 mag sources varies
from 18 030 to 26 935 per deg$^2$
depending on the locations of the bulge fields.  
To search for 2MASS counterparts, we assumed a main association 
radius ($\rm r_{a}$) of 4.0\arcsec.
The density limit and main association radius are chosen such 
as to restrict the chance association to less
than 10\% (where chance association, $\rm y = n \pi r_{a}^{2} \la 0.1$).  
As a result, $\sim$ 89\% of the MSX D-band sources 
(2265 sources with quality flags 3 and 4) within the 2MASS  
observations have an association with a 2MASS source 
($K_{\rm s}$ $\la$ 11.0 mag). Also, 2\% (48 sources) of the MSX D-band sources 
have a 2MASS counterpart which are saturated ($K_{\rm s}$
$\la$ 3.5 mag). 
In Fig.\,\,\ref{FIG2}, we show the 3-D plots of the MSX source
density, the mean A$_{V}$ and the total number of 2MASS sources ($K_{\rm s}$
$\la$ 11.0 mag) in the various subfields of the bulge as a function of the
Galactic longitude and latitude. This information is also
presented in tabular form in Table 1\footnote{Table 1 is only available in the 
electronic form via the VizieR Service at the CDS.}. 

Keeping in mind the importance of high mass-loss AGB stars
in our study, we introduce two additional samples here.
We search for 2MASS counterparts of bright MSX sources detected 
in the D-band but not
associated with 2MASS sources with $K_{\rm s}$ $\la$ 11.0 mag. We
restrict our sample to sources brighter than 4th magnitude in the
MSX D-band. This magnitude limit was chosen such as to include all the
sources with high mass-loss rates ($\rm log_{10}(\dot{M})(M_{\rm
\odot}\,
yr^{-1})$ $\ga -6.0$; see \S 7). There are 165 such sources
with good quality MSX D-band photometry. For identifying 2MASS
counterparts in this additional sample, we extend the main 
association radius to a conservative value of $\sim$ 5\arcsec~without 
any $K_{\rm s}$-band magnitude cut. The increase in the association 
radius by 1\arcsec~ enables us to include some known high mass-loss 
peculiar stars (e.g.  carbon stars) in the sample (see \S 7).
However, a crucial point is that the removal of the $K_{\rm
s}$$\la$ 11.0 magnitude cut increases the 2MASS source
density to $\sim 10^{5}$ sources per deg$^{2}$. Hence,
for a main association radius of 5\arcsec, the chance
association becomes very large ($\rm y \approx 0.6$) 
compared to the 10\% limit set for the rest of the sources in the bulge 
fields. The corresponding pure Poisson chance association is given by 
$\rm 1 - exp(-y) \approx 0.45$. 
This implies that about half of the 2MASS
associations are likely to be spurious. However, it should
be kept in mind that all such spurious associations would
possibly have a $K_{\rm s}$-band counterpart fainter than the
2MASS limit and hence in our study these sources are
retained as a lower limit for $K_{\rm s}$-band and limit
for the high mass-loss end (see \S 6). We are now left with
two samples of these bright MSX D-band sources - 1) Sources
having a 2MASS counterpart within 5\arcsec~radius, which we
designate as sample-A and 2) Sources having no 2MASS
counterpart within this radius, which we name as sample-U. In sample-A, we have
125 sources out of which $\sim$ 70\% have a single
2MASS association within the 5\arcsec~radius. Sample-U contains 42 sources, and we
assign a value of  $K_{\rm s}$ = 13.7 (which is the
$K_{\rm s}$-band limit of the sample-A sources) for all
these sources. 

For this study, alongwith the good quality MSX band
data, we have used only good quality 2MASS $J$-, $H$- and $K_{\rm
s}$-band magnitudes with `rd-flag' values between
1 and 3 which generally implies best quality detection,
photometry and astrometry. However, the 2MASS sources 
with `rd-flag' value of 6 in the $K_{\rm s}$-band (which corresponds to a positive
detection with an upper magnitude limit) were also included with good
quality sources (`rd-flag' 1 -- 3) for the determination of mass-loss
rate only (see \S 6).

\begin{figure}
\centering
\resizebox{\hsize}{!}{\includegraphics{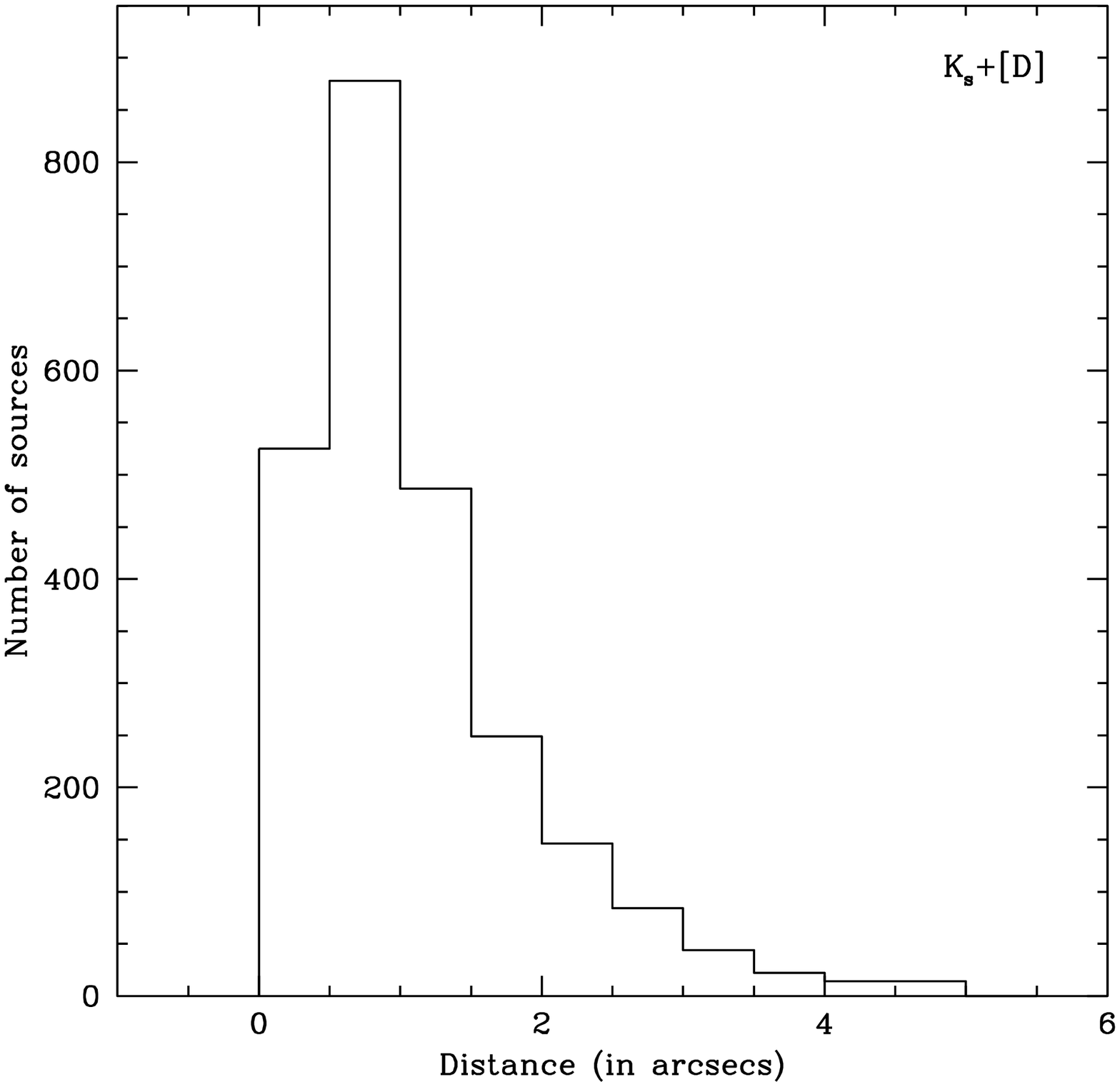}}
\caption{Histogram of positional difference (in arcsec) between good quality 
MSX D-band (quality flags 3 and 4) and 2MASS cross-identified sources.
The MSX--2MASS association was limited to $r$ $<$ 4\arcsec~for
bulk of the sources and increased to 5\arcsec~for sample-A sources (see the
text). 
}
\label{FIG3}
\end{figure}

In Table 2\footnote{Table 2 is available in the electronic form 
via the VizieR Service at the CDS.}, 
we present the full catalogue of MSX--2MASS sources from the bulge fields
(see Fig.\,\,\ref{FIG2} and Table 1), a sample of which is given
in this paper.
\begin{table*}
\setcounter{table}{1}
\caption[]{Sample catalogue of MSX--2MASS sources in the bulge fields.}
\vspace{0cm}
\hspace*{-1cm}
\begin{tabular}{l c c c c c c c c c c c }\\
\hline
Seq. & Name (MSX6C-)  & RA (2000)  & $l$   &   J   &   H   & K$_{\rm s}$$^*$     &  [A]  &  [C]  &  [D]  &  [E]  & A$_{\rm V}$ \\
     &                & Dec (2000) & $b$   &       &       & K$_{\rm s}$$^{**}$  &  [7]  &       &  [15] \\
     &                &   (deg)    & (deg) & (mag) & (mag) &        (mag)        & (mag) & (mag) & (mag) & (mag) & (mag) \\
\hline
6007 & G359.9129+02.1610 & 264.2674 & -0.0871 & 10.11 & 8.16 &  6.99 &  4.15 & 3.13 &  2.96 & 1.99 &  9.48\\
     &                   & -27.8677 &  2.1610 &       &      &  7.11 &  4.72 &  --  & 99.99 &  --  &     \\   
6057 & G359.9338+02.1591 & 264.2818 & -0.0662 &  7.35 & 5.81 &  5.08 &  3.94 & 3.04 &  2.91 & 2.24 &  3.98\\
     &                   & -27.8510 &  2.1591 &       &      &  6.31 &  4.34 &  --  & 99.99 &  --  &     \\  
6444 & G359.9392+02.0232 & 264.4150 & -0.0608 &  6.53 & 5.08 &  4.30 &  3.45 & 2.57 &  2.51 & 2.12 &  3.53\\
     &                   & -27.9193 &  2.0232 &       &      & 99.99 &  3.39 &  --  & 99.99 &  --  &     \\ 
6838 & G000.3366+02.1338 & 264.5491 &  0.3366 &  9.46 & 7.94 &  7.03 &  4.49 & 3.50 &  3.33 & 2.50 &  5.50\\
     &                   & -27.5248 &  2.1338 &       &      &  6.93 & 99.99 &  --  &  3.74 &  --  &     \\ 
6895 & G000.1512+01.9967 & 264.5683 &  0.1512 & 12.05 & 9.72 &  8.12 &  3.95 & 2.75 &  2.56 & 1.70 & 14.59\\
     &                   & -27.7546 &  1.9967 &       &      &  8.33 & 99.99 &  --  &  2.45 &  --  &      \\ 
\hline 
\end{tabular}
\vspace*{0.2cm}
\hspace {-0.5cm}

$^*$~2MASS $K_{\rm s}$ magnitude;~~$^{**}$~DENIS $K_{\rm s}$
magnitude;~~Magnitude assigned
to value of 99.99 means that the source is not detected 

\hspace {-12.0cm}
in the corresponding magnitude band.
\end{table*}
Figure\,\,\ref{FIG3} shows the histogram of positional
difference of MSX D-band--2MASS cross-identified sources in the bulge fields.
The rms of the differences of MSX--2MASS association is
$\sim$ 0.7\arcsec. The histogram peaks at $\sim$1.0\arcsec
~and the almost exponential decreasing trend with increasing distance
indicates good quality associations, which implies that the number of
spurious associations would be minimal in our
cross-correlated sample.

\begin{figure}
\centering
\resizebox{\hsize}{!}{\includegraphics{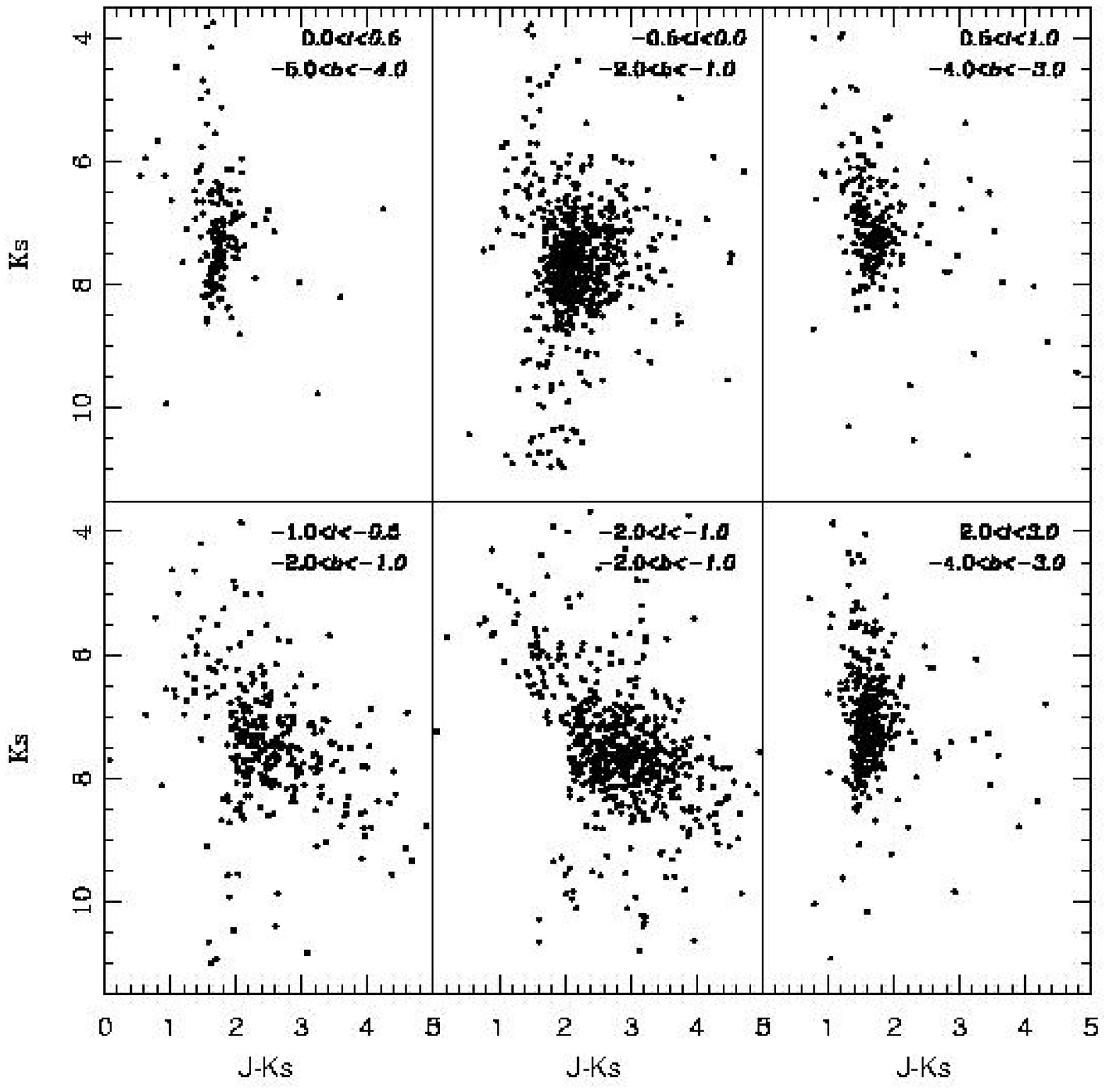}
}
\caption{Color-magnitude diagrams ($J-K_{\rm s}$/$K_{\rm s}$) of 2MASS sources 
in a few selected MSX bulge fields. The CM diagrams for all the bulge
fields will be available in electronic form via the VizieR Service at the CDS. 
} 
\label{FIG4}
\end{figure}

\section{Interstellar extinction and foreground disk stars}

\begin{figure}
\centering
\resizebox{\hsize}{!}{\includegraphics{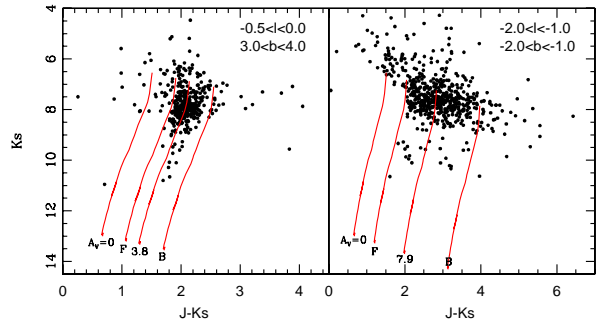}}
\caption{Colour-magnitude diagrams ($J-K_{\rm s}$/$K_{\rm s}$) of MSX--2MASS 
sources in two MSX bulge fields. The four isochrones (Bertelli et al. 1994), 
placed at 8 kpc distance for a 10 Gyr population with Z=0.02 are shown
in the figure for $A_V$ = 0 mag, for the $A_V$ limit adopted for the 
``foreground'' sources (defined as ``F"), for mean $A_V$ 
value of the field and for the $A_V$ limit adopted for the
``B-sources'', respectively. 
} 
\label{FIG5}
\end{figure}

In most parts of the Galactic bulge, the interstellar extinction is not
homogeneous and occurs in clumps, hence, a detailed extinction map is
essential in stellar population studies (Schultheis et al. 1999).
Inspite of the recent 
improvements in the extinction measurements of the 
Galactic bulge (Schultheis et al. 1999; Dutra et al. 2001; Marshall et al. 2006; 
Indebetouw et al. 2005),  
the determinations are still uncertain. In this paper, 
the method as described in Ojha et al. (2003) which is based on the
procedure outlined in Schultheis et al. (1999) is used to determine the
interstellar extinction. The MSX catalogue was divided into 60
subfields. This enables in
decreasing the effects of variable and patchy extinction present in
the entire bulge field. 
For $|l| < 1.0^{\circ}$, we have 0.5 deg$^2$ fields with steps of 0.5
deg in longitude and 1.0 deg in latitude. For rest of the fields,
the field size was increased to 1.0 deg$^2$ with steps of 1.0 deg both in
longitude and latitude. This ensures appreciable statistical
sample in each field to derive mean A$_{V}$. 
The fact that we use 1 deg${^2}$ subfields to derive extinction gives
some dispersion of the order of typically 1.5 mag in $A_{\rm V}$ 
(see Table 1). 
However, at Galactic latitudes greater than 1 deg (where our fields are 
located), the clumpiness of interstellar extinction is less prominent 
(Schultheis et al. 1999).
The subfields are listed in Table 1 and the field centres are
displayed in the various plots of Fig.\,\,\ref{FIG2}. 
Figure\,\,\ref{FIG4}\footnote{
The $J-K_{\rm s}/K_{\rm s}$ colour-magnitude diagrams of 2MASS sources in the 
bulge fields are available in electronic form via the VizieR Service at the 
CDS.} shows the $J-K_{\rm s}/K_{\rm s}$ colour-magnitude
diagrams for a few selected MSX fields presented in
Fig.\,\,\ref{FIG2} and listed in Table 1. 
Figure\,\,\ref{FIG5} shows the $J-K_{\rm s}/K_{\rm s}$ colour-magnitude
diagrams for two sample fields ($-0.5^\circ < l < 0.0^\circ$, 
3$^\circ$ $<$ $b$ $<$ 4$^\circ$ ; $-2.0^\circ < l < -1.0^\circ$, 
-2$^\circ$ $<$ $b$ $<$ -1$^\circ$ ).

As is clearly seen in Figs.\,\,\ref{FIG4} and \ref{FIG5}, the 
$J-K_{\rm s}/K_{\rm s}$ colour-magnitude diagrams of 2MASS sources in the 
bulge fields show a well-defined red giant and AGB sequence shifted by 
fairly uniform extinction, with respect to the reference 
$K_{\rm s0}$ vs. $(J-K_{\rm s})_{\rm 0}$ of Bertelli et al. (1994) with $Z$ = 0.02 
and a distance modulus of 14.5 (distance to the Galactic Centre : 8 kpc). 
However, a point to be noted here is that the intrinsic depth of the 
bulge is about 0.3 mag (Glass et al. 1995).
We have assumed A$_{J}$/A$_{V}$ = 0.256 and A$_{K}$$_{\rm s}$/A$_{V}$ 
= 0.089 (Glass 1999) and calculated the individual extinction values of the
2MASS bulge sources as described in Ojha et al. (2003). 
The mean $A_{V}$ value for each field has been 
determined by a Gaussian fit to the $A_{V}$ distribution.
It should be noted here that owing to low star
counts, we have increased the size of some bins (e.g $0.0 < l <
0.5;\,\, 4.0 < b < 5.0~\&~0.5 < l < 1.0; 4.0 <  b < 5.0$ 
are merged into one bin) to estimate the mean extinction values. 
The details of the subfields, giving
the mean $A_{V}$, the total number of sources and the statistics
of the background and foreground population are listed in 
Table 1. To derive the dereddened magnitudes for the MSX
sources, we have used the extinction law -- A$_{A}$/A$_{V}$ = 0.022; A$_{C}$/A$_{V}$ = 0.023;
A$_{D}$/A$_{V}$ = 0.013; A$_{E}$/A$_{V}$ = 0.016
(Messineo et al. 2002).

As seen in Fig.\,\,\ref{FIG5}, 
the 2MASS sources with anomalously low values of $A_V$ are
probably foreground stars. 
For each field we empirically define an
isochrone ``F" for which we 
assume that all the sources left of it are foreground stars.
Also seen in
Fig.\,\,\ref{FIG5} and in each of the 
bulge fields, these foreground stars are around the isochrone
with $A_{V}$ $\sim$ 0 mag and clearly left of the bulk of the stars 
grouped around the isochrone with mean $A_{V}$ of the
field. These foreground stars will be no
longer considered in the following discussions of bulge
stars.
However, this empirical way of rejecting foreground
population does not exclude foreground stars with significant
mass-loss. 

There are also a number of stars with $J-K_{\rm s}$
values significantly larger than the bulk of the other stars in each bulge
field (right of the isochrone empirically
defined as ``B" 
in Fig.\,\,\ref{FIG5}). The sources to the right
of this isochrone are termed as ``B-sources". We can 
see three reasons for such an excess in $J-K_{\rm s}$ :
1) intrinsic ($J-K_{\rm s}$)$_0$ excess induced by a large mass-loss, which 
should be accompanied by a large 15 $\mu$m excess; 
2) spurious association or wrong photometry which is
rather unlikely for 2MASS sources well above the detection limit.
3) excess in $A_{V}$ which should probably be due to a patchy extinction on the 
bulge line of sight (additional extinction from dust layers behind the Galactic
centre for background stars appears unlikely at such high galactic
latitudes), however this does not seem a major source of excess with relatively 
small average extinction. For the bright
stars in this region, with $K_{\rm s}$ $<$ 8 mag, we use the mean extinction
value (typical $\delta A_{\rm V} \sim 1.56$ mag) of the field. This is
because of two reasons: many of the stars with $K_{\rm s}$ $<$ 8 mag should have a large mass-loss which could produce an excess in ($J-K_{\rm s}$)$_0$; and 
at the bright end we are restricted by the limit of the
Bertelli et al. (1994) isochrone which is used as the reference in
shifting the sources and determining the individual extinctions. 
For rest of the sources in this region which include 
the faint sources, with $K_{\rm s}$ $>$ 8 mag, 
we determine their specific extinction from the $J-K_{\rm s}/K_{\rm s}$
colour-magnitude diagram and the zero-extinction isochrone 
similar to the extinction determination for the bulk of the bulge
sources. For few bright sources in the region left of
isochrone ``B", we have extrapolated the Bertelli isochrone to
estimate the individual extinction values.

The uncertainties in the estimated values of $A_{\rm V}$ due to the
assumption of a 10 Gyr stellar population of solar metallicity as the
reference system has been discussed in Schultheis et al. (1998; 1999).
It is shown that the age-range (of 5 Gyr) has hardly any effect,
whereas, the assumption of solar metallicity results in typical
uncertainties of $\sim$ 1 mag in the determination of $A_{\rm V}$.   
Another important point to be noted here is that for the high mass-loss
AGB stars, post-AGB sources and planetary nebulae population, the
contribution from the intrinsic extinction due to circumstellar dust has not
been considered. This will result in larger uncertainties in the estimated 
$A_{\rm V}$ values. 

The sample of MSX sources discussed in this paper
contains $\sim$ 24000~ 2MASS
sources which are detected in the $K_{\rm s}$-band. From this
sample we have identified ``foreground"
and ``B-sources" based on the method described above
and demonstrated in Fig.\,\,\ref{FIG5}.
Their fractions are about 12\% (2937) and 9\% (2113), respectively. 
For the sample-A population, we have 111 MSX sources (after
foreground subtraction)
which have good quality D magnitudes (D-band quality flags 3 \& 4) and
$K_{\rm s}$-band photometry (`rd-flag' values 1 -- 3). 
It should be kept in mind that for sample-U sources it
is not possible to identify the foreground population and
hence, they are not removed from the sample.

\section{ISOGAL and DENIS association}

We have also cross-correlated all the MSX sources with the ISOGAL sources
in the bulge fields. The ISOGAL Point Sources (IGPSC; Schuller
et al. 2003) have been obtained from the VizieR Service at CDS.
ISOGAL Point Source Catalogue contains about 10$^5$ stars, with associated
$K$, $J$, and $I$ DENIS data for most of them (see Schuller et al. 2003). 
The MSX and ISOGAL associations were searched within a radius of 
6\arcsec~(size of the ISO pixel). There are a total of about 2110 ISOGAL 
sources at 7 and 15\,$\mu$m in the MSX bulge fields 
(total area $\sim$ 48 deg$^2$). Out of this
194 ISOGAL sources at 7 $\mu$m have a MSX
A-band association and 55 ISOGAL sources at 15 $\mu$m 
have MSX D-band association. The corresponding magnitude 
limits for these associated sources are 7.8 and 4.6 mags in the
A- and D-band, respectively. We have
checked the consistency of MSX and ISOGAL 15 $\mu$m magnitudes for the 38
strongest sources (with [D] $<$ 4.0 mag; among our sample-A). The average
difference is [15]$_{MSX}$ - [15]$_{ISO}$ = -0.10$\pm$0.43 mag, where the large
dispersion is probably related to the source variability.
We have searched the DENIS counterparts for only the ISOGAL
cross-correlated MSX sources. More than 99\% of the ISOGAL
sources in the bulge fields have a DENIS counterpart.
There are 435, 61 and 54 2MASS $K_{\rm s}$-band sources which have 
associations with DENIS $K_{\rm s}$ sources, DENIS $K_{\rm s}$ \&
MSX D-band sources and DENIS $K_{\rm s}$, MSX D-band \& ISOGAL 
15 $\mu$m sources, respectively.
For the ISOGAL cross-correlated sources, we have
complemented the 2MASS and MSX sources in our catalogue (Table 2) 
by the DENIS and ISOGAL data.
Within the ISOGAL fields (Ojha et al. 2003),
more than 95\% of MSX sources having 2MASS association have an
ISOGAL counterpart. We have used the DENIS and ISOGAL data,
wherever available, in the estimates of the bolometric magnitudes
(see \S 5) and the mass-loss rates (see \S
6) in order to mitigate the effects of variability. 
Hence, for the bulge sources with DENIS
and ISOGAL associations, we have used the average of the 
2MASS and DENIS values for the $K_{\rm s}$-band magnitude and the
MSX and ISOGAL 15\,$\mu$m values for the D-band magnitudes.

\section{Bolometric magnitudes and luminosities}
\begin{figure}
\centering
\resizebox{\hsize}{!}{\includegraphics{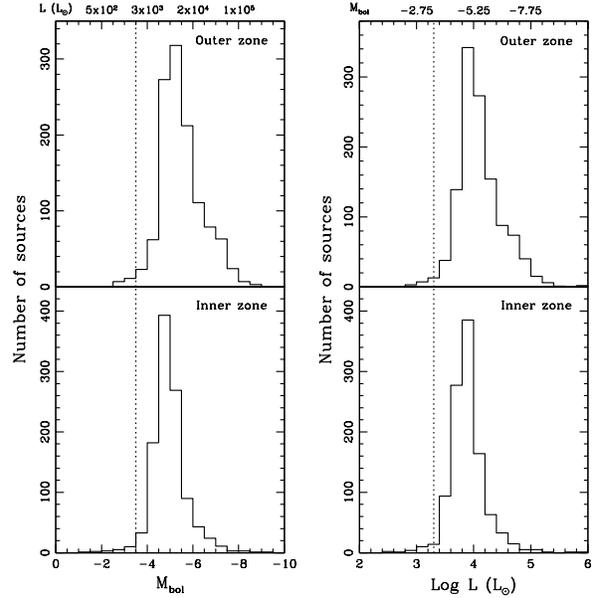}}
\caption{Histograms showing bolometric magnitudes and luminosities of the 
sources in MSX bulge fields which have good quality (quality flags 3 and 4) 
D-band magnitudes and $K_{\rm s}$-band photometry.
The dotted vertical lines indicate the approximate bulge 
RGB tip ($M_{\rm bol}$ $\sim$ -3.5; $L \simeq 2000 L_{\rm \odot}$).}
\label{FIG6}
\end{figure}

After dereddening (see \S 3), the bolometric magnitudes ($M_{\rm bol}$) 
are derived by integrating the flux densities over the wavelength range
between 1.25 $\mu$m $<$ $\lambda$ $<$ 15 $\mu$m by fitting a blackbody
curve. For each object we used the dereddened $J$-, $K_{\rm s}$-, A-, D-band 
magnitudes and the estimated A$_{\rm V}$ values as input parameters.  
The main error of the bolometric corrections results 
from the interstellar extinction values which gives an error 
of $\sim$ 0.2 - 0.3 mag in M$_{\rm bol}$.
The assumption of a blackbody curve for the AGB spectra is less than
optimal given the presence of strong molecular bands. We have
compared the results obtained from the blackbody fitting with that derived
using multi-band bolometric correction for AGB stars (Cecile Loup; private
communication). However, this method is valid for limited range of
colours and luminosities.
The difference in M$_{\rm bol}$ derived for from
the two methods peaks around $\sim$ 0.1 mag which is not that significant.
As mentioned in the previous section, we have used the average of
DENIS \& 2MASS $K_{\rm s}$-band and MSX D-band \& ISOGAL 15 $\mu$m magnitudes 
when available. This helps in reducing the uncertainties in 
M$_{\rm bol}$ and $\rm \dot{M}$
determinations by about a factor $\sqrt{2}$. 
At the same time, it is to be noted that 
large amplitude variable stars such 
as Miras do show variations in the $K_{\rm s}$-band 
 upto $\sim$ 2.5 mag (Groenewegen 2006; Glass et al. 2001).
However, the amplitude of intensity variation depends on the period of these
sources and is typically in the range 0.5 -- 1.0 mag. Hence, using only 
single-epoch measurements give us larger errors
in the bolometric magnitude determination. 
The uncertainties due to this variability can only be reduced by
obtaining time-consuming light curve measurements and this is beyond
the scope of this present work.
It is also important to note
that there are 239 ISOGAL 15 $\mu$m sources which have not been detected
in the MSX D-band (15 $\mu$m) in our sample. These sources have 
also been used in our
sample to determine the M$_{\rm bol}$ and $\rm \dot{M}$. 
The $M_{\rm bol}$ and luminosity ($\rm L (L_{\odot}) =
10^{-(M_{bol} - 4.75)/2.5}$) values of each source are given in
Table 3\footnote{Catalogue of MSX sources from the bulge fields with 
$\rm \dot{M}\, >\, 3\times10^{-7}\, M_{\odot}\, yr^{-1}$ is available in 
the electronic form via VizieR Service at the CDS.}, a sample of which is given in this paper.
Figure\,\,\ref{FIG6} shows the histograms of bolometric magnitudes and 
luminosities of the sources in MSX bulge fields. 
The approximate bulge RGB tip is predicted to be at
$\rm K_{0} \sim 8.0$ mag (Frogel et al. 1999) which translates
to $M_{\rm bol} \sim -3.5$; $\rm L \approx 2000\,
L_{\odot}$. This implies that the distributions are clearly incomplete 
below the luminosity of the bulge RGB tip.
 
\begin{table*}
\setcounter{table}{2}
\caption[]{Sample catalogue of MSX sources from the bulge fields with
$\rm \dot{M}\,>\, 3\times10^{-7}\, M_{\odot}\, yr^{-1}$.
Data for each source are displayed in two lines, with the MSX standard name
(e.g. MSX6C-G359.4989-03.301) and the IRAS name if any. In order to make
the easy comparison with available ISOGAL 7 and 15 $\mu$m magnitudes,
[7] \& [15], and IRAS flux densities (in Jy) at 12 \& 25 $\mu$m, S12 \& S25,
MSX intensities are given in magnitudes for bands A ($\sim$ 8 $\mu$m) and
D ($\sim$ 15 $\mu$m), and in flux densities (in Jy) for bands C
($\sim$ 12 $\mu$m) and E ($\sim$ 21 $\mu$m).}
\vspace{0cm}
\hspace*{-1cm}
{\scriptsize
\begin{tabular}{l c c c c c c c c c c c c}\\
\hline
Name (MSX6C-)     & $l$  &   J   &  H    & K$_{\rm s}$ & [A] & [D] & msxC & msxE &
A$_{\rm V}$ &  L & $\rm \dot {M}$ \\
IRAS Name     & $b$  &  &   & & [7]  & [15] & S12  & S25 &  & &   & \\
 & (deg) & (mag) & (mag) & (mag) & (mag) & (mag) & (Jy) & (Jy) & (mag) & 10$^4$ L$_{\rm \odot}$ & (M$_{\rm \odot}$/yr)\\
\hline
G000.7554-01.0352 & 0.8  & 11.17 &  9.59 &  8.80 & 3.27 & 1.36 & 4.9 & 5.8 &  6.1 & 0.80 & 7.0$\times$10$^{-6}$\\
17482-2848        & -1.0 &       &       &       &      & 1.46 & 5.8 & 6.6\\
G000.0100-03.0215 & 0.0  & 12.49 & 11.92 &  9.82 & 2.99 & 1.59 & 4.7 & 4.4 &  8.7 & 0.99 & 6.3$\times$10$^{-6}$\\
17543-3027        & -3.0 &       &       &       &      & 2.54 & 3.3 & 3.8\\
G359.9506-02.0090 & -0.0 & 14.62 & 12.21 &  9.38 & 2.90 & 1.34 & 5.3 & 5.6 & 22.6 & 1.71 & 2.9$\times$10$^{-6}$\\
                  & -2.0 &       &       &       &      & 2.16 &     &     &\\
G000.0157+01.6944 & 0.0  & 12.92 & 11.42 & 10.74 & 2.75 & 1.15 & 5.6 & 6.3 &  6.9 & 1.05 & 1.6$\times$10$^{-5}$\\
17359-2800        & 1.7  &       &       &       &      &      & 2.9 & 4.6\\
G359.8576+01.0049 & -0.1 & 13.53 & 11.87 & 11.19 & 3.64 & 1.86 & 2.7 & 4.0 &  8.2 & 0.51 & 1.5$\times$10$^{-5}$\\
17382-2830        & 1.0  &       &       &       & 3.47 & 1.54 & 2.0 & 4.3\\
\hline
\end{tabular}
}
\end{table*}

\section{Mass-loss rate ($\rm \dot{M}$) in the bulge} 
\begin{figure}
\centering
\resizebox{\hsize}{!}{\includegraphics{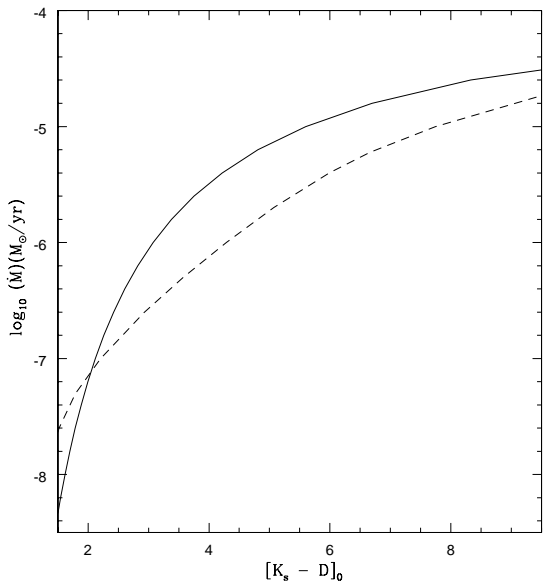}}
\caption{The figure shows the mass-loss rates as a
function of ($K_{\rm s}$-[D])$_{\rm 0}$ colour (see text) for the
Groenewegen (dotted line) and Jeong (solid line) models.}
\label{FIG7}
\end{figure}

We have used the dust radiative transfer models for oxygen-rich AGB stars 
from Groenewegen (2006) to derive the mass-loss rates of the sources in the MSX
bulge fields. These models are simulated for values of $\rm L =
3000\,L_{\odot}$, distance = 8.5 kpc, expansion velocity = 10 km $\rm
s^{-1}$, dust-to-gas ratio = 0.005, and no interstellar reddening. 
Our derivations are based on the model for an oxygen-rich AGB star 
with $\rm T_{eff}$ of 2500 K and 100\% silicate composition is used.
We have used the empirical relation between $\rm \dot{M}$ and 
($K_{\rm s}$-[LW3])$_{\rm 0}$ 
to determine the mass-loss rate of each star. The luminosity distribution of
the sources in our MSX bulge fields peaks at $\sim$ 8000 $L_{\rm \odot}$ (see
Fig.\,\,\ref{FIG6}). 
Using the scaling law given in Groenewegen (2006), we have scaled 
the model curve for this peak luminosity.
It should be noted that the empirical values tabulated
by Groenewegen (2006) are for $\rm \dot{M} < 2\times10^{-5}~M_{\odot}\, yr^{-1}$. 
For the high mass-loss end (see Fig. 8) we have extrapolated the above 
empirical relation used by us.
It should also be noted here that the mass-loss rate derivations
rely on the accurate estimates of the extinction and the
corrections applied to the $K_{\rm s}$ magnitudes.
On the other hand, the colour such as ([A]-[D])$_{\rm 0}$, which depends
little on extinction, allows a direct measure of
the mass-loss rates. However, we do not use this colour for
mass-loss estimates primarily because it is less sensitive
especially for the large and small mass-loss rates.
The model based on the synthetic colours of AGB stars by Jeong et al.
(2002) has also been used in literature for derivation of mass-loss rate (Ojha
et al. 2003). We show a comparison of the Groenewegen's model with that of 
Jeong et al. (2002)  which is significantly different.
In Fig.\,\,\ref{FIG7}, we present the mass-loss
rates derived  as a function of 
($K_{\rm s}$-[LW3])$_{\rm 0}$, using the empirical model of Groenewegen (2006) 
and the empirical relation ($\rm \dot{M}$ vs. ($K_{\rm s}$-{\rm 15})$_{\rm 0}$) 
of Jeong et al. (2002). Here, in place of these colours, we use
($K_{\rm s}$-[D])$_{\rm 0}$. This is justified because
the central wavelength of MSX-D band is 14.65\,$\mu$m which is
close to 15\,$\mu$m and falls within the LW3/ISO filter (12 --
18\,$\mu$m).  
It is seen that for Vega the difference in
magnitudes ($[D] - [15]$) is 0.04 mag which is
negligible. It is to be noticed that in the whole 
range of mass-loss we are interested in, for $\rm \dot{M}\, \ga\,
10^{-7}\, M_{\odot}\, yr^{-1}$, the curve based on Groenewegen's
model gives lower values of mass-loss by factors 
$\ga$ 3 as compared to the scale of Jeong et al. (2002) for
similar colour indices. 
In this paper, we present the mass-loss rates derived from the models of
Groenewegen (2006). 
The two mass-loss rate scales are also presented later in
the discussions of the nature of the MSX bulge sources (see Figs.\,\,\ref{FIG11} 
and \ref{FIG12}).

\begin{figure}[h]
\centering
\resizebox{\hsize}{!}{\includegraphics{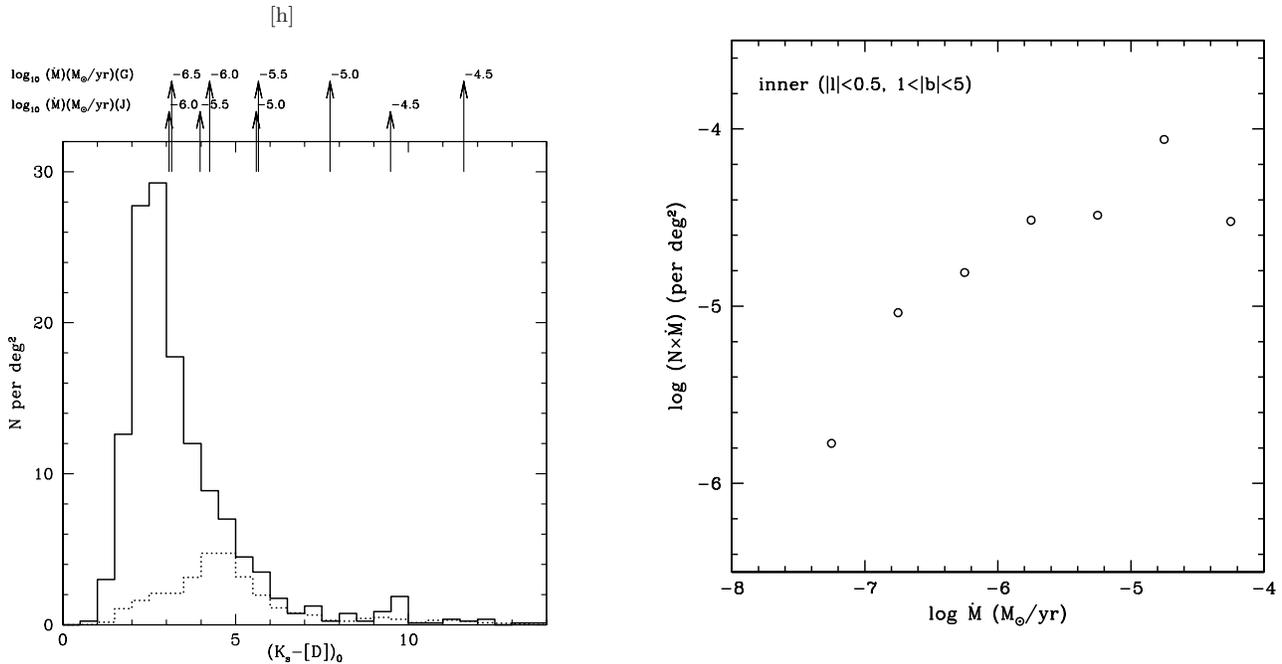}}
\caption{$K_{\rm s}$-[D] colour distribution of MSX sources. The mass-loss 
rate scales displayed at the top of the figure are from the models of
Groenewegen (upper panel) and Jeong (lower panel). The 
full line histogram represents the stars of the ``inner" zone 
($|l| < 0.5^\circ$) and the dashed line denotes stars of the
``outer" zone  
($0.5^\circ < |l| < 3.0^\circ$) in the bulge (Fig.\,\,\ref{FIG2}
and Table 1). 
}
\label{FIG8}
\end{figure}

Fig.\,\,\ref{FIG8} shows the $K_{\rm s}$-[D]$_{\rm 0}$ 
colour distribution of MSX sources in the ``inner" and ``outer" zones in the bulge
fields. There are certain issues regarding the
mass-loss estimates which need to be mentioned. It should be noted 
that using the empirical model of Groenewegen (2006)  
includes uncertainties in the determination of the mass-loss rates arising from 
the following reasons: (1) The relation was derived for oxygen-rich AGB stars 
in the solar neighbourhood. It has also been argued that metallicity affects 
the dust-to-gas ratio and the outflow velocity from evolved stars
(Habing et al. 1994). This is 
directly related to the mass-loss and, therefore, the colour-mass-loss relation 
could possibly differ in different environments such as between the Galactic 
bulge and the Magellanic Clouds. (2) It is important to emphasize that the
AGB stars are variable in nature.
The average $K$ amplitude of sources associated with known 
LPVs (Glass et al. 2001) is $\sim$ 1.0 mag, which amounts to a factor of
$\sim$ 2 -- 5 uncertainty in the determination of $\rm \dot{M}$.
(3) The model (Groenewegen 2006) used
is valid for a limited range of luminosity and expansion
velocity and are simulated for AGB stars. Groenewegen (2006) have also mentioned
about the caveat of post-AGB model simulation. Here the model for post-AGB
sources is calculated under the assumption that the effective temperature and 
luminosity do not change over the time for the dust shell to drift away.
So using the colour-mass-loss relation for sources having these parameters 
beyond the range (i.e sources with high luminosity and expansion velocity like
PNe and post-AGB or sources with lower luminosity like T-Tauri
stars) will have a large uncertainty in their mass-loss estimation.
The classical AGB luminosity limit is $\rm Log\, L(L_{\odot})\,
\sim\,
4.74$
(Marigo et al. 1998; Zijlstra et al. 1996). Hence, in our MSX bulge
fields, sources having luminosity values beyond this cut-off limit 
will have larger uncertainties in the
mass-loss estimation. These sources and their contribution to the integrated
mass-loss rate will be discussed in detail later in this section.

\begin{figure*}[h]
\includegraphics[width=17cm]{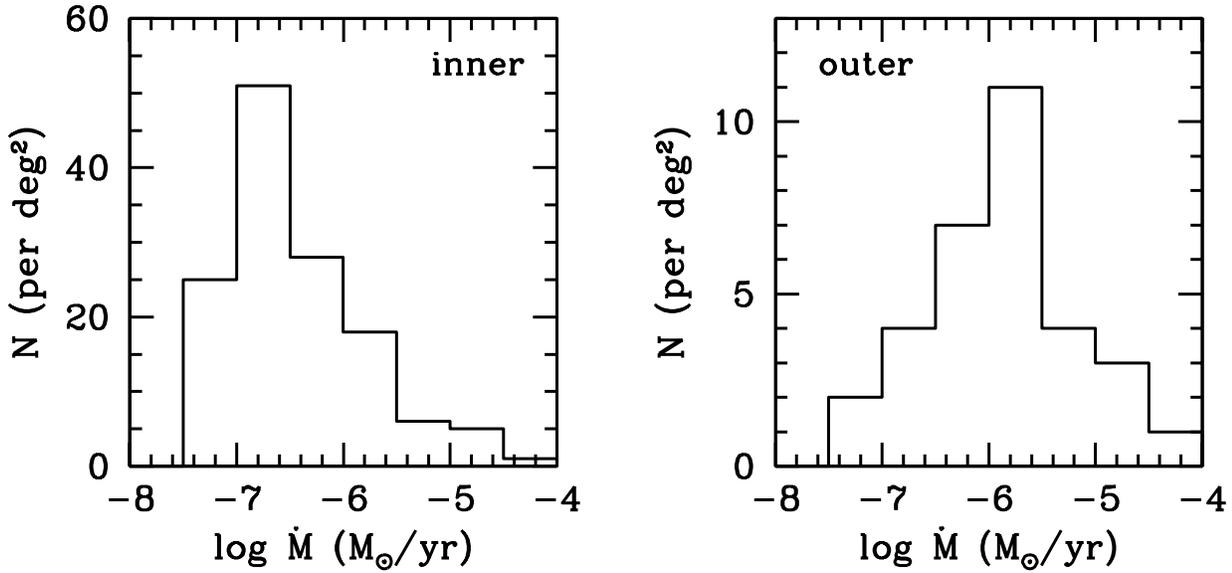}
\caption{Distribution of mass-loss rates ($\rm \dot {M}$) of MSX sources
in the ``inner" (left) and ``outer" (right) bulge zones. The mass-loss rates
are inferred from the {Groenewegen's model} using the
($K_{\rm s}$-[D])$_{\rm 0}$ colour. 
}
\label{FIG9}
\end{figure*}

\begin{figure}
\resizebox{\hsize}{!}{\includegraphics{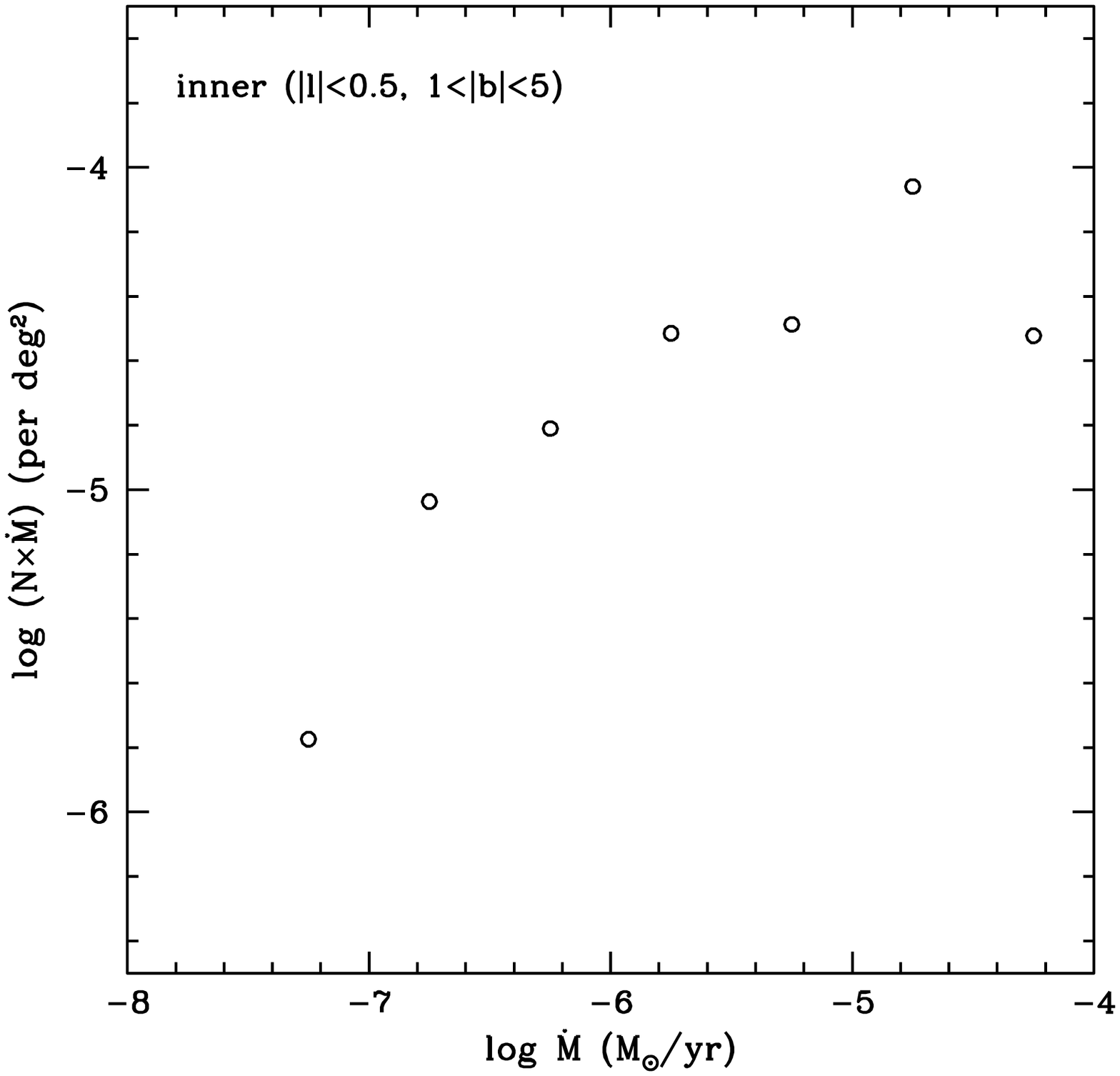}}
\caption{Average total mass-loss rate per square degree and per 0.5 bin of
log ($\rm \dot {M}$) as a function of $\rm \dot {M}$ for MSX sources
in the ``inner" bulge zone.
}
\label{FIG10}
\end{figure}

The mass-loss rates for the objects in the MSX bulge fields range
from $10^{-7}$ to $\rm 10^{-4}\,M_{\odot}\, yr^{-1}$. 
As is obvious from the comparison of the two histograms in 
Fig.\,\,\ref{FIG8}, the number of sources is incomplete for 
$\rm \dot{M}\, \la \,3\times10^{-7}\, M_{\odot}\, yr^{-1}$ 
(Groenewegen's model) in the ``inner" zone, and, more severely, 
for $\rm \dot{M}\, \la\, 2\times10^{-6}\, M_{\odot}\, yr^{-1}$ 
in the ``outer" zone. 
The source excess seen particularly in the ``inner"
zone for $K_{\rm s}$-[D] $\sim$ 9 -- 10, is most likely due to spurious
association of the sample-A and sample-U sources 
(see \S 2), which makes the estimated mass-loss rates appear smaller than their 
actual values.
 Numerical values of the mass-loss rates
of individual sources are given in Table 3.
The number distribution of mass-loss rates in the ``inner" and ``outer" bulge
zones as a function of $\rm \dot{M}$ are displayed in Fig.\,\,\ref{FIG9}. 
%
In Fig.\,\,\ref{FIG10}, we plot the 
corresponding values of the average total mass-loss rate per square degree 
and per 0.5 bin of $\rm log\,\dot{M}$ for the ``inner" zone. We do not 
attempt to
build the same plot for the ``outer" zone because all the mass-loss bins in
this zone, except  $\rm \dot{M} \,\sim\, 0.3 - 3 \times 10^{-5}\,
M_{\odot}\, yr^{-1}$, would be very uncertain. 
This is due to the incompleteness of the lower
mass-loss bins and the uncertainty in $K_{\rm s}$-[D]$_{\rm 0}$ for
 $\rm \dot{M}\, \ga\, 3\times10^{-5}\, M_{\odot}\, yr^{-1}$.
%
Numerical values of the integrated mass-loss
rate in the ``inner" zone are displayed in Table 4, where 
we have limited the integration to mass-loss rates, 
$\rm \dot{M} > 3\times10^{-7}~M_{\odot}$ yr$^{-1}$ for which the 
data are reasonably complete. 
The first row shows the value of mass-loss rate per square 
degree averaged over all the fields in the ``inner" zone.
In the next five rows, we give the integrated mass-loss in the five
latitude bins which cover the bulge zone. The table also lists the 
corresponding integrated
mass-loss rate per unit stellar mass (in $\rm yr^{-1}$).
The integrated mass-loss rate in the entire ``inner" zone for the bulge fields is 
$\rm 1.96\times10^{-4}\, M_{\odot}\, yr^{-1}\, deg^{-2}$ and the
corresponding integrated mass-loss rate per unit stellar mass is 
$\rm 0.48\times10^{-11}\, yr^{-1}$ (see Table 4). 
For the rest of the discussion that follows, it should
be kept in mind that the integrated mass-loss rate is only for the ``inner"
zone and the contribution from the ``outer" zone is not included.

Here, we would like to discuss about the sources with luminosities beyond
the classical AGB limit of  $\rm Log\, L(L_{\odot})\, > \,4.74$. 
There are 18 such sources
in the ``inner" zone for which we derive the integrated mass-loss rate. 
Out of these only three sources have mass-loss rates 
$\rm \dot{M}\, >\, 3\times10^{-7}\, M_{\odot}\, yr^{-1}$. The uncertainty in the
mass-loss estimation of these three sources will not affect the integrated
mass-loss rate as they contribute less than 4\% to the total mass-loss. We have
also examined the nature of all these 18 sources in various CM and CC diagrams.
Except for one source which has a faint $K_{\rm s}$-band magnitude
(12.3 mag), rest of sources are very bright in $K_{\rm s}$ ($<$ 5.6 mags) 
with 4 of them saturated in 2MASS ($K_{\rm s}$ $<$ 3.5 mag). All of these
sources are also bright in the MSX D-band ($<$ 2.8 mag). 
Referring to Figs.\,\,\ref{FIG11} and \ref{FIG13}, we see that 
the majority of these sources have blue colours 
(($K_{\rm s}$-[D])$_{\rm 0}$ $\sim$ 2; ([A]-[D])$_{\rm 0}$ $\sim$ 1). This
implies that they are mostly foreground sources which have possibly
not been removed using our foreground source rejection procedure (see
\S 3). The foreground (or early-type) nature of
these sources were further verified in the $(I-J)/(K_{\rm s}-[D])$ CC diagram
based on the discussion presented in Schultheis et al. (2002; cf Fig. 2). 
Among these 18 sources, there
are two sources which show large colour excess in the diagrams
discussed above. These are likely to be YSOs. Additional 
spectroscopic and photometric observations are required to understand
the nature of these luminous sources. 
However, the uncertainties involved in the mass-loss rates 
of these sources do not influence the estimation of the integrated
mass-loss rate as their contribution is negligible as is already mentioned.

We have also calculated the integrated
mass-loss rate for the ``intermediate" ($|$$l$$|$ $<$ 3.0$^\circ$,
1$^\circ$ $<$ $|$$b$$|$ $<$ 2$^\circ$) 
and the ``outer" ($|$$l$$|$ $<$ 3.0$^\circ$,
2$^\circ$ $<$ $|$$b$$|$ $<$ 5$^\circ$) bulge fields. For  
$\rm \dot{M}\, >\, 3\times10^{-7}\, M_{\odot}\, yr^{-1}$, the integrated
mass-loss rates are $\rm 3.87\times10^{-4}\,M_{\odot}\,yr^{-1}\,deg^{-2}$ 
and $\rm 1.32\times10^{-4}\, M_{\odot}\,yr^{-1}\,deg^{-2}$ in the 
two bulge fields, respectively.
In comparison, Ojha et al. (2003) derive values of 
$\rm 3.7\times10^{-4}\, M_{\odot}\,yr^{-1}\,deg^{-2}$ and 
$\rm 0.4\, -\, 1.0\times10^{-11}\, yr^{-1}$ for the integrated mass-loss 
rate and  mass-loss rate per unit stellar mass, respectively,
using ISOCAM 7 and 15 $\mu$m observations. It should be
noted here that the 
Galactic bulge fields presented in Ojha et al. (2003)
cover $-1.5^\circ<l<+1.6^\circ$; $-3.8^\circ<b<-1.0^\circ$, $b=+1.0^\circ$ 
with a total area of $\sim$ 0.29 deg$^{2}$ whereas
the area covered in this
present work is much larger and hence offers better statistics than the
sample of Ojha et al. (2003). The
sensitivities of the MSX and 2MASS data also allow us to probe the high
mass-loss end with a better statistical sample.

Le Bertre et al. (2001; 2003) have studied the Galactic mass-losing 
AGB stars and concluded that the replenishment
to the ISM is dominated ($\ga$ 50\%) by AGB stars with mass-loss rates
$\rm \ga\, 10^{-6}\, M_{\odot}\,yr^{-1}$. These sources
constitute $\sim$10\% of their sample and noticeably
there are no AGB stars in their sample with mass-loss rates 
$\rm > \,10^{-4}\, M_{\odot}\,yr^{-1}$ though  sources with 
mass-loss $\rm \ga\,10^{-4}\, M_{\odot}\,yr^{-1}$
are known to exist (Habing 1996). 
The sensitivities of the data
sets used by us probe these rare high mass-loss stars. In our MSX
bulge fields, we have a total of 42 sources displaying mass-loss 
$\rm \ga\, 1.0\times10^{-5}\, M_{\odot}\,yr^{-1}$ which significantly 
contribute to the mass returned to the ISM. In comparison,
the stellar population studies in the solar neighbourhood by Jura
\& Kleinmann (1989) identify 63 sources with mass-loss rates 
$\rm \ga \,10^{-6}\, M_{\odot}\,yr^{-1}$ out of which only 
21 sources in their sample have
mass-loss rates $\rm \ga\, 1.0\times10^{-5}\, M_{\odot}\,yr^{-1}$. 
From their sample it is seen that oxygen- and carbon-rich AGB stars have almost equal
contribution to the replenishment of the ISM which amounts to 
$\rm 3\,-\, 6\times10^{-4}\, M_{\odot}\,kpc^{-2}\,yr^{-1}$. 
From the MSX bulge fields we estimate the total mass replenishment to the ISM 
to be $\rm \sim \,9\times10^{-3}\, M_{\odot}\,kpc^{-2}\,yr^{-1}$,
which is not surprising given the large number of high mass-loss
AGB stars ($\rm \dot{M}\,\ga\, 1.0\times10^{-5}\,M_{\odot}\,yr^{-1}$)
detected in our sample. 
However, it should also be kept in mind that the sample
of Jura \& Kleinmann (1989) is incomplete at both the
lower and the high mass-loss ends.
Since the low mass-loss
end is incomplete in our sample, we refrain from making
any comparison with other studies.
\begin{table}
\setcounter{table}{3}
\centering
\caption[]{Integrated mass-loss (in $\rm M_{\odot}\, yr^{-1}$
deg$^{-2}$) and integrated mass-loss rate per unit stellar mass 
(in $\rm yr^{-1}$) in the ``inner'' bulge zone with 
$\rm \dot{M}\, >\, 3\times10^{-7}\,M_{\odot}\,yr^{-1}$. 
The mass-loss rates are calculated using the model of
Groenewegen (2006).}
\begin{tabular}{c c c c c c c}\\
\hline
Inner Fields & $\rm M_{\odot}\, yr^{-1}$ deg$^{-2}$  & yr$^{-1}$ \\
\hline
All & 1.96$\times$10$^{-4}$ & 0.48$\times$10$^{-11}$  \\
\hline
1.0$^\circ$ $<$ $|$$b$$|$ $<$ 1.5$^\circ$ & 4.34$\times$10$^{-4}$ & 0.52$\times$10$^{-11}$ \\
1.5$^\circ$ $<$ $|$$b$$|$ $<$ 2.0$^\circ$ & 3.41$\times$10$^{-4}$ & 0.54$\times$10$^{-11}$  \\
2.0$^\circ$ $<$ $|$$b$$|$ $<$ 3.0$^\circ$ & 1.66$\times$10$^{-4}$ & 0.35$\times$10$^{-11}$ \\
3.0$^\circ$ $<$ $|$$b$$|$ $<$ 4.0$^\circ$ & 1.50$\times$10$^{-4}$ & 0.41$\times$10$^{-11}$  \\
4.0$^\circ$ $<$ $|$$b$$|$ $<$ 5.0$^\circ$ & 0.80$\times$10$^{-4}$ & 0.27$\times$10$^{-11}$ \\
\hline
\end{tabular}
\end{table}

\section{Nature of the MSX sources}

In this section, we discuss the various colour-magnitude and
colour-colour diagrams. This enables us to study the nature
of the sources in the MSX bulge fields. As has been mentioned earlier,
we present only sources with good quality MSX and 2MASS data.
 
\subsection{Colour-magnitude diagrams}

\begin{figure}
\centering
\resizebox{\hsize}{!}{\includegraphics{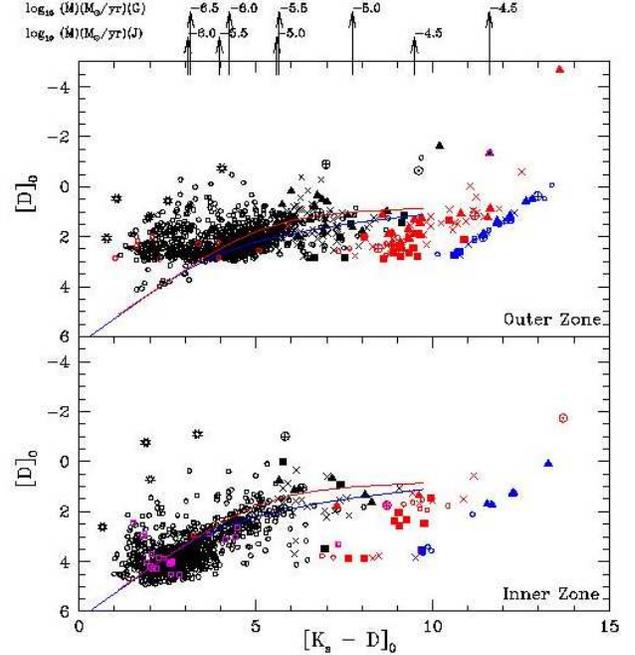}}
\caption{[D]$_{\rm 0}$/($K_{\rm s}$-[D])$_{\rm 0}$ magnitude-colour diagram 
for the sources in the MSX bulge fields. The upper panel shows the
``outer"
zone ($0.5^\circ<|l|<3.0^\circ$) and the lower panel shows
the ``inner" zone ($|l|<0.5^\circ$) of the bulge. Only good quality data
(MSX flags 3 \& 4; 2MASS `rd-flg' 1 -- 3) are presented.
The colour coding which is followed is red for sample-A
sources, blue for sample-U sources and black for rest of the bulge fields.
The mass-loss rate scales displayed on top are from 
Groenewegen (2006) and Jeong et al. (2002) (lower) models.
The two curves are from the models of 
Groenewegen (2006), where the red curve represents the
model for an oxygen-rich AGB 
star with $\rm T_{eff}$ = 2500 K and a 100\% silicate composition. The blue 
curve is for a carbon-rich AGB star with $\rm T_{eff}$ = 2650 K with 100\% 
amorphous carbon (AMC). 
The model curves are scaled to the peak luminosity of $\rm
8000\,L_{\odot}$ of the sources in the bulge fields.
In case of the Groenewegen models, the magnitude in the 
ISOCAM LW3 (15\,$\mu$m) filter has been used in place of 
the MSX D-Band.
The ISOGAL counterparts are marked as magenta open squares for the
entire bulge fields including the sample-A sources.
The known sources from the VizieR and SIMBAD database are 
displayed with the following symbols while maintaining the colour
coding for the sample-A, sample-U and the rest of the bulge fields 
sources : Planetary Nebulae and Post AGB stars - Solid squares; Maser and OH/IR
sources - solid triangles; Peculiar sources including variable stars -
$\oplus$; IRAS sources with no other classifications - crosses.
The two carbon stars in our sample are represented by $\odot$.
Sources saturated in the 2MASS $K_{\rm s}$-band ($\le 3.5$) are
shown as stars.}
\label{FIG11}
\end{figure}

\begin{figure}
\centering
\resizebox{\hsize}{!}{\includegraphics{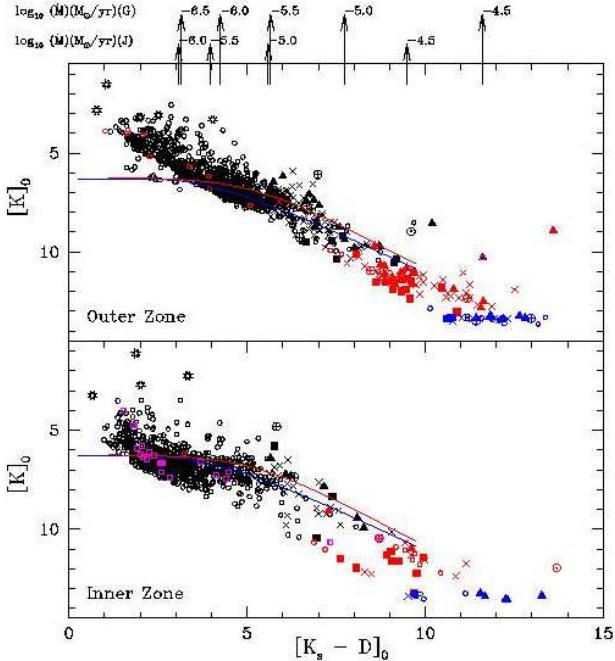}}
\caption{$K_{\rm s0}$/($K_{\rm s}$-[D])$_0$ magnitude-colour diagram of MSX
sources with 2MASS counterparts. The symbols are same as shown in 
Fig.\,\,\ref{FIG11}. 
}
\label{FIG12}
\end{figure}

\begin{figure}
\centering
\resizebox{\hsize}{!}{\includegraphics{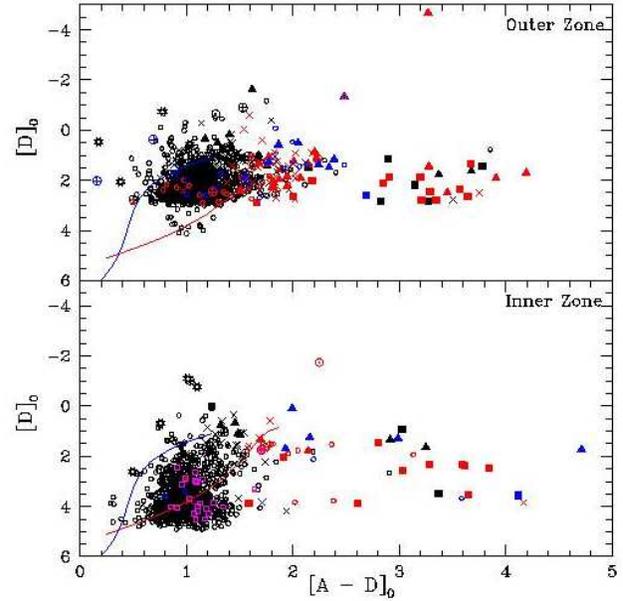}}
\caption{[D]$_{\rm 0}$/([A]-[D])$_{\rm 0}$ magnitude-colour diagram of 
good quality MSX sources. The IRAC 8\,$\mu$m magnitude is used in place of MSX
A-band for the Groenewegen models. The symbols are same as shown in
Fig.\,\,\ref{FIG11}.}
\label{FIG13}
\end{figure}

\begin{figure}
\centering
\resizebox{\hsize}{!}{\includegraphics{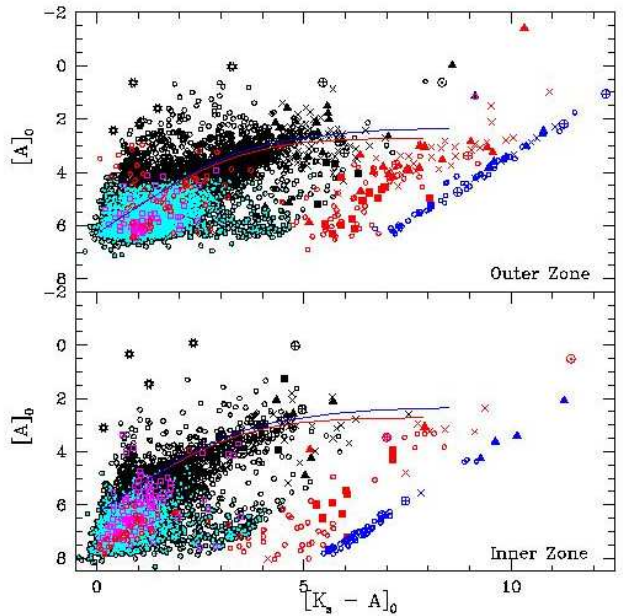}}
\caption{[A]$_0$/($K_{\rm s}$-[A])$_0$ magnitude-colour diagram of MSX sources
(quality flags of 3 \& 4) with 2MASS counterparts. The upper panel shows the 
``outer" zone ($0.5^\circ < |l| < 3.0^\circ$) and the lower panel shows the 
``inner" zone ($|l| < 0.5^\circ$) of the bulge. 
The symbols are same as in Fig.\,\,\ref{FIG11}. The cyan
circles represent the sources without D associations.
}
\label{FIG14}
\end{figure}

In Figs.\,\,\ref{FIG11}, \ref{FIG12}, \ref{FIG13} and
\ref{FIG14}, we present the dereddened ($K_{\rm s}$-[D])$_{\rm 0}$/[D]$_{\rm 0}$,
($K_{\rm s}$-[D])$_{\rm 0}$/$K_{\rm s}$$\rm {_0}$, 
([A]-[D])$_{\rm 0}$/[D]$_{\rm 0}$ and 
($K_{\rm s}$-[A])$_{\rm 0}$/[A]$_{\rm 0}$
colour-magnitude diagrams of MSX sources in the two bulge zones,
respectively.

Figures\,\,\ref{FIG11} and \ref{FIG12} show the two
colour-magnitude diagrams involving the ($K_{\rm
s}$-[D])$_{\rm 0}$ colour which has been used in \S 6 for the
derivation of the mass-loss rates of the sources. 
The mass-loss rate scales derived from Jeong et al. (2002) and Groenewegen (2006)
are displayed on the top panel of these two figures. It is interesting to see
that the sources from sample-A (which are plotted in red colour) are mostly 
high mass-loss stars with 
($K_{\rm s}$-[D]$)_{\rm 0}$ $\ga$ 6.5, corresponding to a mass-loss rate
$>$ $\rm 5\times 10^{-6}\,M_{\odot}\, yr^{-1}$ based on the Groenewegen's
scale.
The sample-U sources (shown in blue) extend the high mass-loss 
sequence beyond
the sample-A sources. It should be kept in mind that the straight
line sequence seen for the sample-U sources is partly due to the
assigned lower limit of 13.7 mag on the $K_{\rm s}$-band magnitude.
It is worth recalling here that these two samples constitute sources
brighter than 4th mag in MSX D-band with no 2MASS counterpart with $K_{\rm
s}$ $\la$ 11.0 mag within 4\arcsec~search radius. Sample-A includes sources 
having a 2MASS counterpart within an extended 5\arcsec~search radius without
any  $K_{\rm s}$-band magnitude cut and sample-U comprises of sources 
having no 2MASS counterpart within this radius. We have assigned
$K_{\rm s}$ = 13.7 mag for all the sample-U sources. 
Another distinct feature which is seen in the ``inner" zone is branching
of the sources into two sequences with an appreciable gap
around $K_{\rm s}$-[D]$)_{\rm 0} \ga 5.5$. 
It is also interesting to note that majority of the ISOGAL
sources seen in the ``inner" bulge zone, i.e. lower luminosity sources, 
occupy the low mass-loss end of the plots.
Our prime focus in this study is to identify the nature
of high mass-loss stars in the bulge fields. We therefore selected
sources with large mass-loss rates ($> \rm
3\times10^{-5}~M_{\odot}yr^{-1}$ as per the Groenewegen's scale)
and surveyed the VizieR and SIMBAD database for identification 
with known objects. Since majority of these high mass-loss sources 
are known IRAS sources, we used a search radius of 10\arcsec~ by 
taking into consideration the typical IRAS PSC error ellipse 
of $\sim$ 10\arcsec $\times$ 20\arcsec. We have grouped
these sources identified in VizieR/Simbad into four major classes -- 
Planetary nebulae \& post AGB stars, maser \& OH/IR sources, peculiar sources
(which also include variable stars) and IRAS sources which
only have IRAS names but no other classifications. We have
marked these four classes in the colour-magnitude
diagrams using different symbols (see caption of
Fig.\,\,\ref{FIG11}).

In Fig.\,\,\ref{FIG13}, we have plotted the
([A]-[D])$_{\rm 0}$/[D]$_{\rm 0}$ colour-magnitude diagram. 
There is a gap seen around ([A]-[D])$_{\rm 0}$ $\sim$ 2.5. Among
the sources to the right of this gap (the redder sources) the
majority are known post AGB and planetary nebulae.
In this figure it is also clearly evident that sources
from sample-A are redder than the rest. 
Figure\,\,\ref{FIG14} shows 
($K_{\rm s}$-[A])$_{\rm 0}$/[A]$_{\rm 0}$ colour-magnitude
diagram. The ($K_{\rm s}$-[A])$_{\rm 0}$/[A]$_{\rm 0}$ diagram seems to be the 
best criterion for the detection of large amplitude LPVs (Glass et al. 1999,
Schultheis et al. 2000). We see two interesting sequences in this
colour-magnitude diagram. The sources seem to branch out
into two for ($K_{\rm s}$-[A])$_{\rm 0}$ $\ga$ 2, with an appreciable gap
between the two. The upper sequence primarily consists of
sources with $J-K$ excess and the lower sequence mostly comprises of
sources which do not have any D-band association. Interestingly,
the majority of the sources in the lower sequence do not
have either C- or E-band association and are fainter in $K_{\rm s}$
($>$ 9 mag). 
The sources from sample-A seem to extend the sequence
with increasing slope of the curve with increasing ($K_{\rm s}$-[A])$_{\rm 0}$ 
colour and the majority of these are again known post AGB and planetary nebulae. 
The sample-U sources seem to form a parallel sequence and occupy the
red end. As has been mentioned earlier, it must be noted that this sequence 
is also a result of the assigned lower limit of 13.7 mag on the $K_{\rm s}$-band
magnitude of the sources in this sample.
In both the above diagrams the ISOGAL sources are
towards the blue end. 

\subsection{Colour-colour diagrams}

\begin{figure}
\centering
\resizebox{\hsize}{!}{\includegraphics{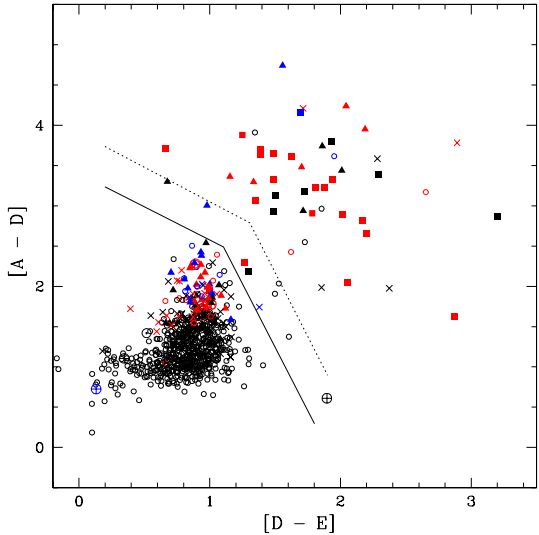}}
\caption{($D-E$)/($A-D$) colour-colour diagram for the two bulge
zones are presented (open black circles). The known objects from the VizieR and 
SIMBAD database are displayed. The symbols are same as shown in
Fig.\,\,\ref{FIG11}. 
The dotted line from Ortiz et al. (2005) shows the position of the gap between. 
We have added the solid line which allows one to identify a few additional
border-line objects}
\label{FIG15}
\end{figure}

The colour-colour diagrams are also useful for discriminating
between different classes of objects and to study their nature.
Ortiz et al. (2005), have studied the evolution of
carbon- and oxygen-rich AGB stars, post-AGB stars and
planetary nebulae using mainly data from MSX.
In comparison to their Fig. 3, we have
plotted the ($D-E$)/($A-D$) colour-colour diagram in
Fig.\,\,\ref{FIG15}. We have plotted the reddened colours to
facilitate comparison with the plot of Ortiz et al. (2005).
As noted by Ortiz et al. (2005), we
also find a distinct gap in the plot which is shown as a 
solid line based on the visual inspection of our data, in 
addition to the line (dotted one) from Ortiz et al. (2005). 
The solid line based on our estimate allows to discriminate few
additional border-line objects.
In this figure, we have plotted the known high
mass-loss sources, which have already been identified.
Apart from these we have also searched
the VizieR and SIMBAD database for known counterparts of all the
additional
sources lying to the right of the gap, which were missed out
in the earlier selection based on Fig.\,\,\ref{FIG11}, which
included only good quality $K_{\rm s}$-band sources.
Majority of the known
IRAS and maser sources lie to the left and lower end of the
plot. Planetary
nebulae populate mostly the region beyond the gap. Ortiz et
al. (2005) suggest that the gap seen is due to the rapid
evolution of stars as they cross the dotted line 
which can be considered as a transition boundary. This is in good agreement
with the different population of sources found on either
side of the gap in Fig.\,\,\ref{FIG15}. 

\begin{figure*}
\includegraphics[width=6.3cm]{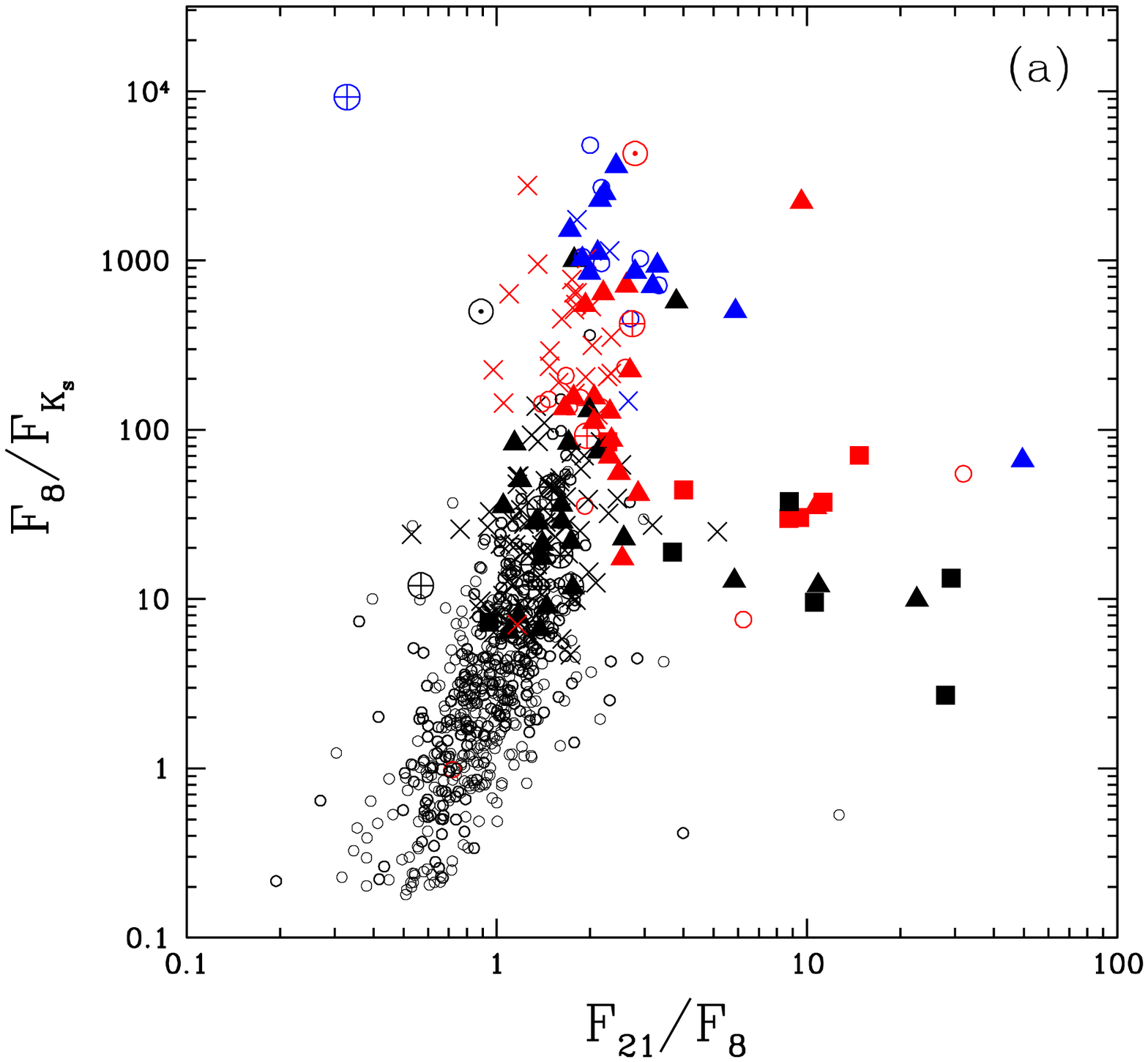}
\qquad
\includegraphics[width=6.3cm]{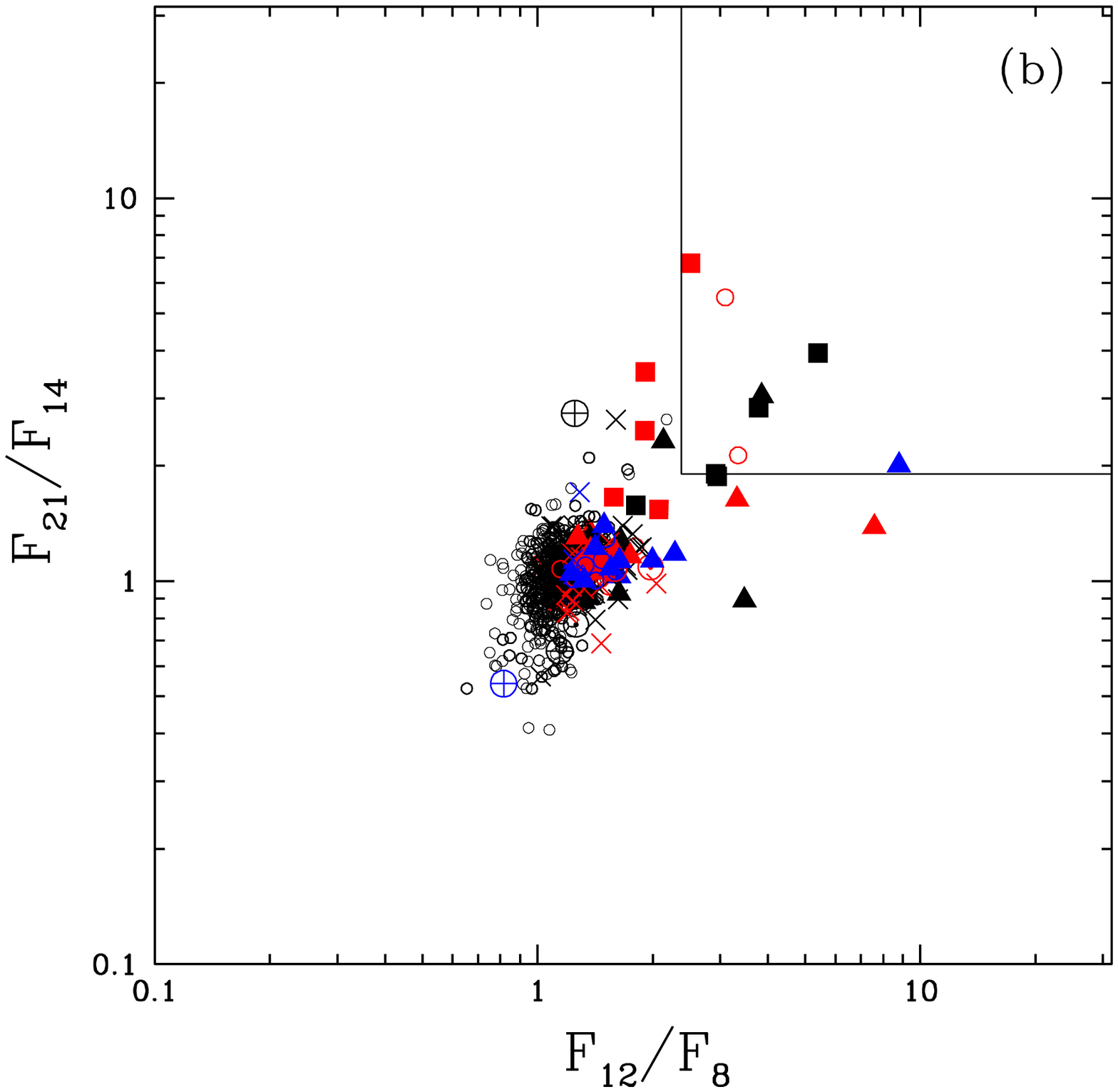}
\includegraphics[width=6.3cm]{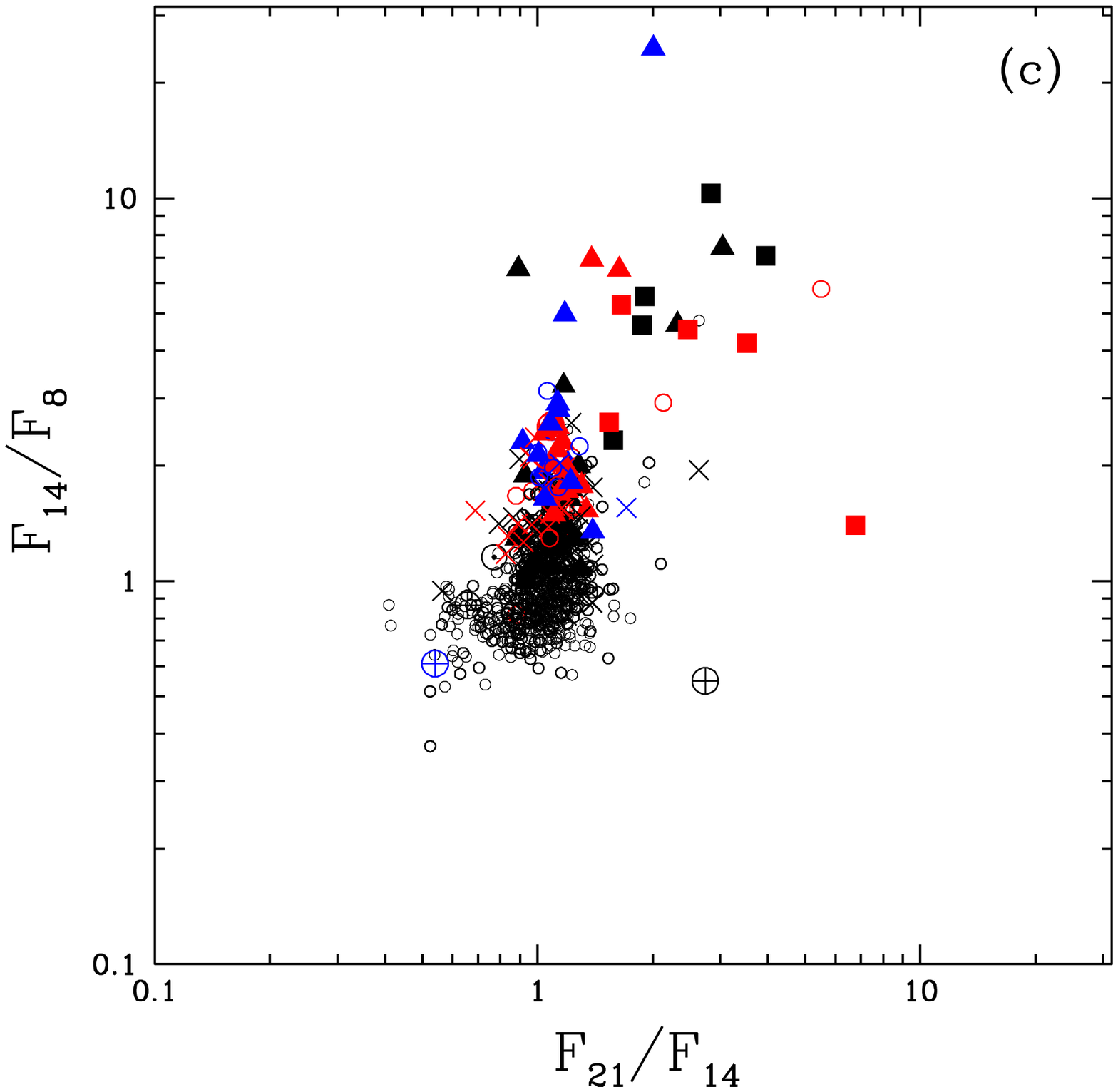}
\qquad
\includegraphics[width=6.3cm]{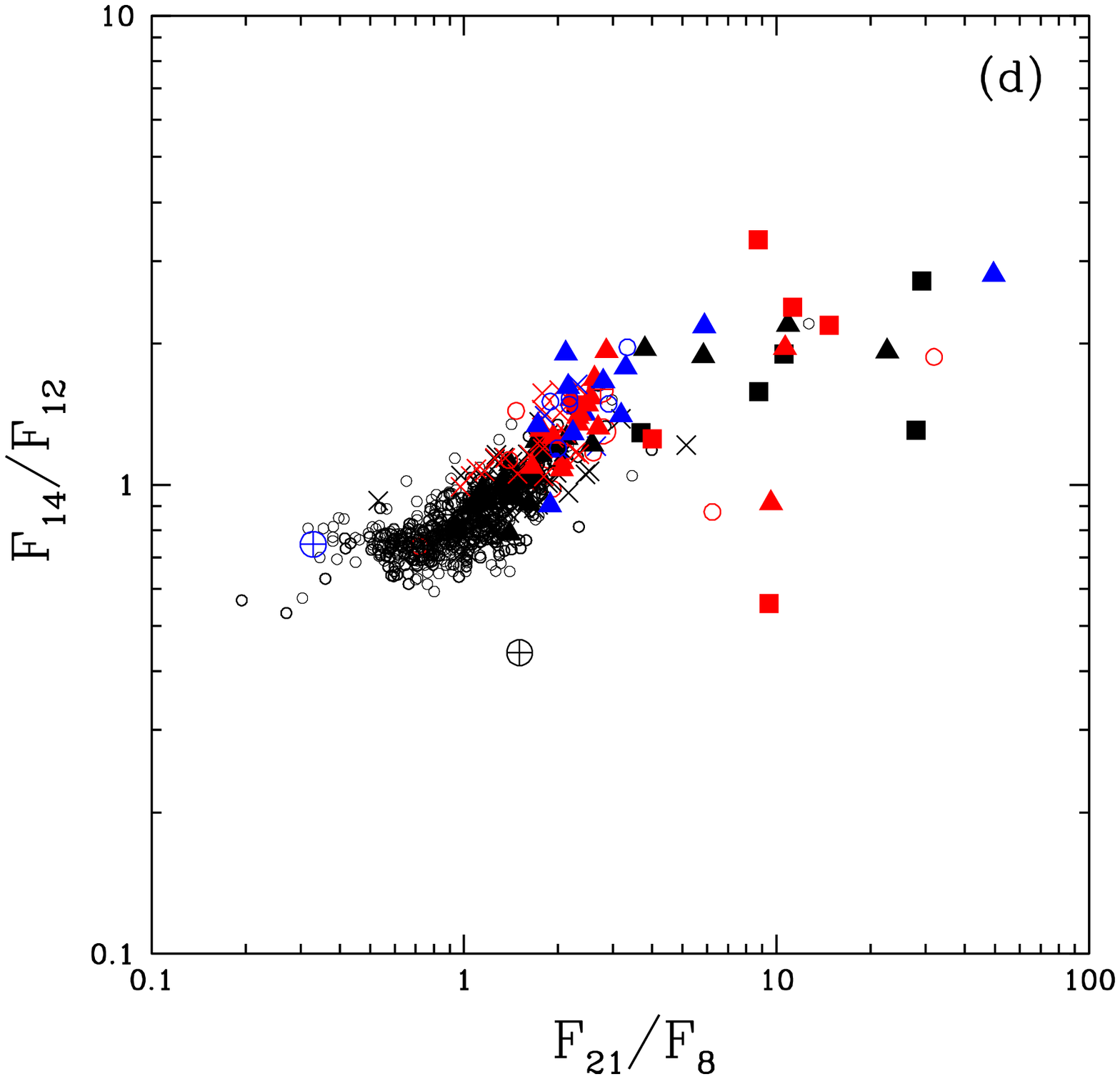}
\caption{The near- and mid-infrared colour-colour diagrams for the two
bulge zones (open black circles). The rectangular box shown in (b) is based on the
criteria discussed in Messineo et al. (2004).
The symbols 
are same as shown in Fig.\,\,\ref{FIG11}.}
\label{FIG16}
\end{figure*}

Using MSX and 2MASS mid- and near-infrared colour-colour diagrams, 
Lumsden et al. (2002), Sevenster (2002) and Messineo et al.
(2004) have studied different classes of evolved stars with
circumstellar envelopes. In
Fig.\,\,\ref{FIG16}, we present the different colour-colour diagrams of
the MSX bulge fields for good quality flags in all the MSX bands and
also in the $K_{\rm s}$-band. The plots are shown in the
units of the reddened flux ratio rather than the colour index for
easy comparison with the plots of Lumsden et al. (2002). 
In Fig.\,\,\ref{FIG16}a, we plot the 
$\rm F_{21}/F_{8}$ vs. $\rm F_{8}/F_{K}$ flux ratios. According to Lumsden
et al. (2002), the oxygen- and carbon-rich populations separate out
clearly in this colour-colour diagram 
(see Figs. 10 \& 11 of Lumsden et al. 2002). However,
we do not see any such separation here in our MSX bulge fields. 
The OH/IR stars and the planetary
nebulae mostly have $\rm F_{21}/F_{8}$ flux ratios $\ga$ 1 consistent
with the trend seen in Fig. 10 of Lumsden et al. (2002).
The mid-infrared colours of planetary nebulae are similar to HII
regions but mostly planetary nebulae are bluer because they are not
embedded in molecular clouds.
Figure\,\,\ref{FIG16}b shows the 
$\rm F_{12}/F_{8}$ vs. $\rm F_{21}/F_{14}$ colour-colour diagram.  
This diagram is useful to distinguish between the AGB and the post-AGB phases and 
to locate post-AGB stars which have redder colours due to their 
colder envelopes (Sevenster 2002; Messineo et al. 2004).
The transition from a blue ($<$ 2.38) to red ($>$ 2.38) $\rm F_{12}/F_{8}$ flux 
ratio may correspond to a transition off the AGB to 
proto-planetary nebulae and from a blue ($<$ 1.90) to red ($>$ 1.90) 
$\rm F_{21}/F_{14}$ flux ratio indicates
a later evolutionary transition, when mass-loss starts to drop down 
several order of magnitudes and there is the onset of the fast wind 
(Messineo et al. 2004). Most of the MSX sources show
$\rm F_{12}/F_{8}$ $<$ 2.38 and $\rm F_{21}/F_{14}$ $<$ 1.90 
as expected for AGB stars. 
In Fig.\,\,\ref{FIG16}c, we plot the $\rm F_{21}/F_{14}$ vs. $\rm
F_{14}/F_{8}$ colour-colour diagram. We do not see any trend separating
the oxygen-rich population from the carbon-rich one as pointed out by
Lumsden et al. (2002). The majority of the sources are
OH/IR stars and IRAS sources which occupy the central region of
the plot. The planetary nebulae are mostly above $\rm
F_{21}/F_{14} \ga 1 $  and $\rm F_{14}/F_{8} \ga 1$. We see a gap around
$\rm F_{14}/F_{8} \sim 3$. This could possibly suggest the transition
from the AGB to the more evolved phase.
In Fig.\,\,\ref{FIG16}d, we plot the $\rm F_{21}/F_{8}$ vs. $\rm
F_{14}/F_{12}$ flux ratios. The MSX sources with 
$\rm F_{21}/F_{8}$ $\ga$ 5.96 (see also
Fig.\,\,\ref{FIG16}a) probably represent
post-AGB stars (Messineo et al. 2004).

\subsection{Population of carbon stars in the bulge fields}

\begin{figure}
\centering
\resizebox{\hsize}{!}{\includegraphics{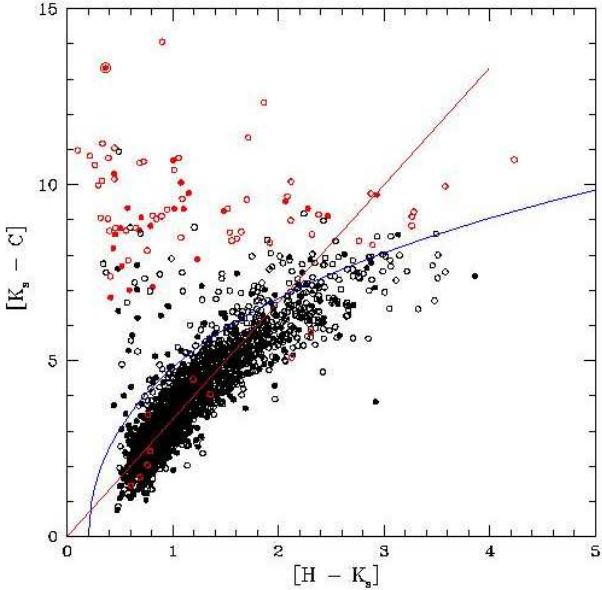}}
\caption{($H-K_{\rm s}$)/($K_{\rm s}$-[C]) colour-colour diagram for
the ``inner" (solid circles) and ``outer" (open circles) zones are
presented. The sources from sample-A are shown in red. The
carbon-rich (red line) and oxygen-rich (blue curve)
sequences from van Loon et al. (1998) are also plotted. One of
the two carbon stars identified in our sample is marked with $\odot$. 
}
\label{FIG17}
\end{figure}

\begin{figure}
\centering
\resizebox{\hsize}{!}{\includegraphics{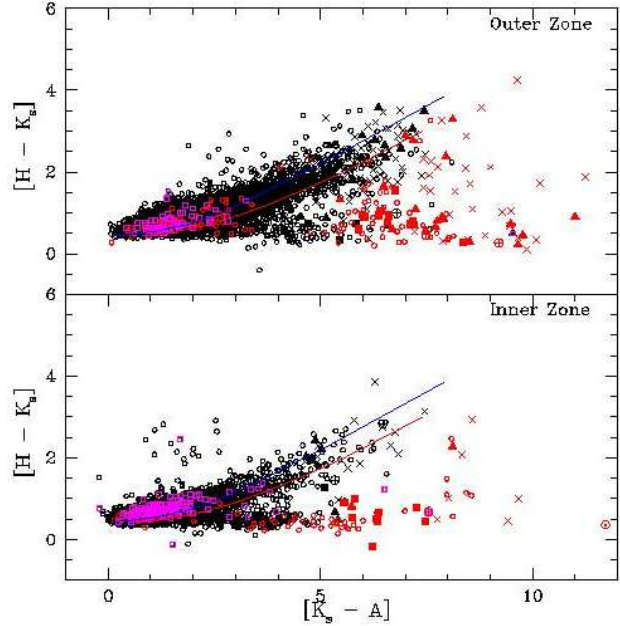}}
\caption{($H-K_{\rm s}$)/($K_{\rm s}$-[A]) colour-colour diagram for
the ``inner" (solid circles) and ``outer" (open circles) zones are
presented. The symbols are same as shown in 
Fig.\,\,\ref{FIG11}.
}
\label{FIG18}
\end{figure}

We have investigated the nature of the red sources
(with $J-K_{\rm s} > 3$) in the various diagrams presented. They occupy the high
mass-loss end of Figs.\,\,\ref{FIG11} and \ref{FIG12}
as expected. From the loci of the Groenewegen's models, 
it is difficult to
comment on whether these sources are carbon-rich or oxygen-rich AGB
stars. The population of carbon-rich stars if
present in the bulge would be crucial in our
understanding of the bulge formation, star
formation history in the bulge and the nature,
formation and evolution of this rare population itself.
Except for the peculiar 34 carbon stars detected by Azzopardi
et al. (1991), there has been no other carbon star detection
in the bulge. The bulge membership and nature of these stars have been
a matter of much debate (Ng 1998; Whitelock 1993). 
In Figs. \,\,\ref{FIG11} and \ref{FIG12}, we see
sources falling on the locus of the carbon-rich AGB star model.
But based on just the colour and magnitude, it is impossible to
confirm or reject the possibility of finding carbon stars in
the bulge. 
Apart from observations focussed on the Galactic bulge, there are 
several other studies based on the near- and mid-infrared colour-colour
and colour-magnitude diagrams to identify carbon star populations in 
different Galactic environments (van Loon et al. (1998); 
Buchanan et al. (2006); Thorndike et al. (2007); Groenewegen et al. (2007)).
van Loon et al. (1998) have presented the
($H-K$) vs ($K-[12]$) colour-colour diagram to distinguish
between carbon and oxygen rich AGB stars. The Figs. 3a and b
of van Loon et al. (1998) show the Galactic
sample of AGB stars from Guglielmo et al. (1993). In their
figures, the oxygen
and carbon rich sequences clearly separate out. In
Fig.\,\,\ref{FIG17}, we have plotted the similar colour-colour
diagram (($H-K_{\rm s}$)/($K_{\rm s}-C$)) of the sources in our bulge fields. 
In this plot we have plotted only sources from sample-A and rest of
the bulge fields. Sources from sample-U, which do not have $H$-band
magnitudes, are excluded.
To compare with the figures of van Loon et al. (1998), we have plotted
here the reddened colours. The scatter in our plot makes
it difficult to differentiate the two sequences. 
In their study of luminous sources in the LMC, Thorndike et al.
(2007) have used the (($H-K$)/($K-[A]$)) colour-colour diagram
to classify different stellar populations. 
In Fig.\,\,\ref{FIG18}, we have plotted a similar reddened
(($H-K_{\rm s}$)/($K_{\rm s}-A$)) colour-colour diagram for the sources in the
MSX bulge fields. 
Here also we see the branching off sources into two sequences similar to that
seen in Fig.\,\,\ref{FIG14}.
The oxygen- and carbon- rich 
AGB star models of Groenewegen (2006) trace the sources in the 
upper sequence better. 
However, given the scatter in the plot, it is difficult to identify
any carbon-rich AGB star sequence as shown in Thorndike et al. (2007).
We also derive similar conclusions from the investigation of the 
(($J-K_{\rm s}$)/($K_{\rm s}-[A]$)) colour-colour
diagram (not presented in this paper) based on the plots presented 
in Buchanan et al. (2006) and Groenewegen et al. (2007). 
However, it is interesting to note that there are two known carbon stars
(IRAS 17534-3030 and IRAS 17547-3249) in our
sample, one each in the ``inner" and the ``outer" zone. 
These sources classified
as carbon stars (Groenewegen et al. 2002; Volk et al. 2000; Guandalini et
al. 2006) do not seem to fall on the carbon-rich model curve shown in  
Figs.\,\,\ref{FIG11}, \ref{FIG12}, \ref{FIG13} and
\ref{FIG14}. One of the two carbon stars (IRAS
17534-3030), which lies in the ``inner" zone of
sample-A, has good quality 2MASS $H$, $K_{\rm s}$, MSX A- and C-band
data and is shown in Fig.\,\,\ref{FIG17} and Fig.\,\,\ref{FIG18}. 
It is seen to lie far away from the carbon line. 
Here, we would like to recall that for sample-A sources, the chance
associations are large ($\approx 0.6$; see \S 2) and hence the possibility of 
a spurious 2MASS association. However, for this particular 
source which is from sample-A, we have checked the near- and
mid-infrared colours in detail. Comparing the colours with the plots
presented in Lumsden et al. (2002), we see that the flux ratios agree
well with the carbon star population except for the $\rm F_{21}/F_{8}$ 
and $\rm F_{21}/F_{12}$ ratios which are marginally on the redder
side. It is worth mentioning here that this carbon star IRAS
17534-3030 is a well known extreme carbon star which has been
recently studied by Pitman et al. (2006) using ISO-SWS spectra. They
predict the presence of nitride dust in the star.
Apart from this, only near-infrared colour-colour diagrams have
also been used to identify carbon stars. Based on the study of 
Kerschbaum \& Hron (1994), we have also checked the ($J-H$) vs ($H-K{\rm s}$)
colour-colour diagram (not presented in this paper). Here also, we are
unable to identify any carbon star sequence except for the
identified carbon star IRAS 17534-3030 whose ($J-H$)/($H-K{\rm s}$)
colours are consistent with the locus of carbon stars.
Lebzelter et al. (2002) and Cioni \&
Habing (2003), have shown the use of the
($I-J$) vs ($J-K$) colour-colour diagram to
discriminate between the oxygen- and carbon-rich
populations for the LMC. Since most of the sources (including the 
two known carbon stars) in our sample do not have $I$ counterparts, 
we have not presented this colour-colour diagram in this paper. 
From the figure of
Ortiz et al. (2005), the carbon stars are seen to populate
the lower left corner (bluer end -- [14.7]-[21.3] colour ranging from $\sim$
0.2 to 0.8 and [8.3]-[14.7] colour ranging from $\sim$ 0 to 1.5) of the
diagram. One of the two carbon stars in our sample falls in
this region (see Fig.\,\,\ref{FIG15}).
Other than this, there is no such distinct population 
seen among the sources belonging to our bulge fields. 
We infer that based on only
the colour-colour and colour magnitude diagrams we cannot conclusively
comment on the presence or absence of carbon stars in the bulge.

\section{Conclusion}

In this paper, we have presented in detail the study of the AGB
population detected by MSX in the ``intermediate" ($|$$l$$|$ $<$ 3.0$^\circ$,
1$^\circ$ $<$ $|$$b$$|$ $<$ 2$^\circ$)  and ``outer"  ($|$$l$$|$ $<$ 3.0$^\circ$, 2$^\circ$ $<$ $|$$b$$|$ $<$ 5$^\circ$) Galactic bulge fields 
covering a large area of 48 deg$^{2}$. We have shown that the
MSX data in conjunction with the 2MASS database can be potentially used to
detect AGB stars well above the RGB tip with the determination of their
luminosities (provided they belong to the bulge) and mass-loss rates.
The sensitivities of the two surveys enable us to sample the high 
mass-loss end, characterized by excess in $K_{\rm s}$-[D])$_{\rm 0}$ colour
due to circumstellar dust emission, with better
statistics. We have derived the mass-loss rates of all the MSX sources
which range from $10^{-7}$ to $\rm 10^{-4}\, M_{\odot}\, yr^{-1}$ and 
estimated the
integrated mass-loss rate in the MSX bulge fields. Taking the
completeness into account, we have limited the integration to
mass-loss rates, $\rm \dot{M} > 3\times10^{-7}~M_{\odot}$ yr$^{-1}$ for 
the ``inner" zone ($|l| < 0.5^\circ$) only. There is a factor
of $\sim$ 3 increase in the estimated integrated mass-loss rate
in the ``intermediate" ($\rm 3.87\times10^{-4}\,
M_{\odot}\, deg^{-2}\,yr ^{-1}$) as compared to the ``outer"
($\rm 1.32\times10^{-4}\, M_{\odot}\, deg^{-2}\,yr ^{-1}$) bulge
regions. The average integrated mass-loss rate is estimated to be 
$\rm 1.96\times10^{-4}\, M_{\odot}\, deg^{-2}\,yr ^{-1}$ 
and the corresponding integrated mass-loss rate per
unit stellar mass is $\rm 0.48\times10^{-11}\, yr^{-1}$.
Apart from the mass-loss derivations, we have used the
various colour-magnitude and colour-colour diagrams to discuss in detail
the nature of the MSX sources in the bulge fields and identify the
location of planetary nebulae, post-AGB stars, OH/IR sources etc. in these
diagrams.

Studies based on proposed {\it Spitzer} IRAC and MIPS photometric surveys of the innermost
Galaxy including the most obscured and crowded region would in
future enable us to unravel in detail the late stages
of stellar evolution particularly of the infrared-luminous AGB stars, the
mass-loss of which is crucial in understanding the later stellar fate and
hence the dust in the universe. Future infrared mission (e.g.
WISE ; Duval et al. 2004) hold the key to such studies.

\section{Acknowledgements}
We thank the anonymous referee for the comments and
suggestions that helped in improving the paper.
This publication makes use of data products from the Two Micron All Sky
Survey, which is a joint project of the University of Massachusetts and the
Infrared Processing and Analysis Center/California Institute of Technology,
funded by the National Aeronautics and Space Administration and the National
Science Foundation.

This research made use of data products from the Midcourse Space Experiment,
the processing of which was funded by the Ballistic Missile Defence
Organization with additional support from NASA office of Space Science.
We would like to thank C. Loup for providing the
program to calculate the bolometric magnitudes.


\begin{thebibliography}{}
\bibitem{}Azzopardi M., Rebeirot E., Lequeux J., Westerlund, B. E., 1991, A\&AS, 88, 265
\bibitem{}Beichman C. A., Chester T. J., Skrutskie M., Low F. J., Gillet F., 1998, PASP, 110, 480
\bibitem{}Bertelli G., Bressan A., Chiosi C., Fagotto F., Nasi E., 1994, A\&AS, 106, 275
\bibitem{} Buchanan C. L., Kastner J. H., Forrest W. J., Hrivnak
B. J., Sahai R., Egan M., Fran A., Barnbaum C., 2006, ApJ, 132, 1890
\bibitem{}Cioni M.-R. L., Habing H. J, 2003, A\&A, 402, 51
\bibitem{}Dutra C. M., Bica E., Clari\'{a} J. J., Piatti A. E., Ahumada A. V., 2001, A\&A, 371, 895
\bibitem{}Duval V. G., Irace W. R., Mainzer A. K., Wright E. L., 2004,
Proceedings of the SPIE, Vol. 5487, page 101
\bibitem{}Egan M. P., Price S. D., Moshir M. M., Cohen M., Tedesco E., 1999, {\it MSX Point Source Catalog Explanatory Guide Version 1.2}, AFRL-VS-TR-1999-1522, Air Force Research Laboratory, AD-A381933
\bibitem{}Egan M. P., Price S. D., Kraemer K. E., Mizuno D. R., Carey S. J., Wright C. O., Engelke C. W., Cohen M., Gugliotti M. G., 2003, {\it MSX Point Source Catalog Explanatory Guide Version 2.3}, AFRL-VS-TR-2003-1589, Air Force Research Laboratory, AD-A381933 
\bibitem{}Epchtein N., de. Batz B., Copet E., Fouque P., Lacombe F., Le Bertre T., Mamon G., Rouan D., et al., 1994, ApSS, 217, 3
\bibitem{}Frogel J. A., Tiede G. P., Kuchinski L. E., 1999, AJ, 117, 2296
\bibitem{}Glass I. S., Whitelock P. A., Catchpole R. M., Feast M. W., 1995, MNRAS, 273, 383
\bibitem{}Glass I. S., 1999, in ``Handbook of Infrared Astronomy'',
Cambridge University Press
\bibitem{}Glass I. S., Ganesh S., Alard C., Blommaert J. A. D. L., Gilmore G., Lloyd Evans T., Omont A., Schultheis M., Simon G., 1999, MNRAS, 308, 127
\bibitem{}Glass I. S., Matsumoto S., Carter B. S., Sekiguchi K., 2001, MNRAS, 321, 77
\bibitem{}Glass I. S., Schultheis M., 2002, MNRAS, 337, 519
\bibitem{}Groenewegen M. A. T., 1997, in ``The Impact of Large Scale Near-IR Sky Surveys'', F. Garzon et al. (eds). Kluwer. Page 165
\bibitem{}Groenewegen M. A. T., Sevenster M., Spoon H. W. W., P\'{e}rez I., 2002, A\&A, 390, 511
\bibitem{}Groenewegen M. A. T., Blommaert J. A. D. L., 2005, A\&A, 443, 143
\bibitem{}Groenewegen M. A. T., 2006, A\&A, 448, 181
\bibitem{} Groenewegen M. A. T., Wood P. R., Sloan G. C., et al.
2007, MNRAS, 376, 313
\bibitem{}Guandalini R., Busso M., Ciprini S., Silvestro G., Persi P., 2006, A\&A, 445, 1069
\bibitem{}Guglielmo F., Epchtein N., Le Bertre T., et al., 1993, A\&AS, 99, 31
\bibitem{}Habing H. J., Tignon J., Tielens A. G. G. M., 1994, A\&A, 286, 523
\bibitem{}Habing H. J., 1996, A\&AR, 7, 97
\bibitem{}Indebetouw R., Mathis J. S., Babler B. L., Meade M. R., Watson C., Whitney B. A., Wolff M. J., Wolfire M. G., et al., 2005, ApJ, 619, 931
\bibitem{}Jeong K. S., Winters J. M., Le Berte T., Sedlmayr E., 2002, in Mass-losing Pulsating Stars and their Circimstellar Matter, ed. Y. Nakada, M. Honma \& M. Sekin
\bibitem{}Jura M., Kleinmann S. G., 1989, ApJ, 341, 359
\bibitem{} Kerschbaum F., Hron J., 1994, A\&AS, 106, 397
\bibitem{}Le Bertre T., Matsuura M., Winters J.M., Murakami H., Yamamura I., Freund M., Tanaka M., 2001, A\&A, 376, 997
\bibitem{}Le Bertre T., Tanaka M., Yamamura I., Murakami H., 2003, A\&A, 403, 943
\bibitem{}Lebzelter T., Schultheis M., Melchior, A. L., 2002, A\&A, 393, 573
\bibitem{}Lumsden S. L., Hoare M. G., Oudmaijer R. D., Richards D., 2002, MNRAS, 336, 621
\bibitem{}Marigo P., Bressan A., Chiosi C., 1998, A\&A, 331, 564
\bibitem{}Marshall D. J., Robin A. C., Reyl\'{e} C., Schultheis M., Picaud S., 2006, A\&A, 453, 635
\bibitem{}Messineo M., Habing H. J., Sjouwerman L. O., Omont A., Menten K. M., 2002, A\&A, 393, 115
\bibitem{}Messineo M., Habing H. J., Menten K. M., Omont A., Sjouwerman L. O., 2004, A\&A, 418, 103 
\bibitem{}Ng Y. K., 1998, astro-ph/9808067
\bibitem{}Ojha D. K., Omont A., Schuller F., Simon G., Ganesh S., Schultheis M., 2003, A\&A, 403, 141 
\bibitem{}Omont A., Ganesh S., Alard C., Blommaert J. A. D. L., Caillaud B., et al., 1999, A\&A, 348, 755
\bibitem{}Omont A., Gilmore G., Alard C., Aracil B., August T., et al., 2003, A\&A, 403, 975
\bibitem{}Ortiz R., Blommaert J. A. D. L., Copet E., 2002, A\&A, 388, 279
\bibitem{}Ortiz R., Lorenz-Marins S., Maciel W. J., Rangel, E. M., 2005, A\&A, 431, 565
\bibitem{}Pitman K. M., Speck, A. K., Hofmeister, A. M., 2006, MNRAS, 371, 1744
\bibitem{}Price S. D., Egan M. P., Carey S. J., Mizuno D. R., Kuchar, T. A., 2001, AJ, 121, 2819
\bibitem{}Schuller F., Ganesh S., Messineo M., Moneti A., Blommaert J. A. D. L., Alard C., et al., 2003, A\&A, 403, 955
\bibitem{}Schultheis M., Ng Y. K., Hron J., Kerschbaum F., 1998, A\&A,
338, 581
\bibitem{}Schultheis M., Ganesh S., Simon G., Omont A., Alard C., Borsenberger J., Copet E., Epchtein N., Fouqué P., Habing H., 1999, A\&A, 349, 69
\bibitem{}Schultheis M., Ganesh, S., Glass, I.S., et al. 2000, A\&A, 362, 215
\bibitem{}Schultheis M., Glass I. S., 2001, MNRAS, 327, 1193
\bibitem{}Schultheis M., Parthasarathy M., Omont A., Cohen M., Ganesh S.,
Sevre F., Simon G., 2002, A\&A, 386, 899  
\bibitem{}Schultheis M., Lan\c{c}on A., Omont A., Schuller F., Ojha D. K., 2003, A\&A, 405, 531
\bibitem{}Sedlmayer E., 1994, in Molecules in the Stellar Environment, ed. U. G. Jorgensen (Berlin: Springer), 163
\bibitem{}Sevenster M. N., 2002, AJ, 123, 2772 
\bibitem{}Skrutskie M. F., Cutri R. M., Stiening R., 2006, AJ, 131, 1163
\bibitem{} Thorndike S. L., Kastner J. H., Buchanan C., Hrivnak
B. J., Sahai R., Egan M., 2007, {\it astro-ph/0703584}
\bibitem{}van Loon J. Th., Zijlstra A., Whitelock, P. A., et al., 1998, A\&A, 329, 169
\bibitem{}Volk K., Xiong G-Z., Kwok S., 2000, ApJ, 530, 408
\bibitem{}Whitelock P., 1993, IAUS, 153, 39 
\bibitem{} Zijlstra A. A., Loup C., Waters L. B. F. M., Whitelock P., van Loon J. Th., Guglielmo F., 1996, MNRAS, 279, 32
\end{thebibliography}
\end{document}